\documentclass[12pt]{article}
\usepackage{lineno}

\usepackage{fullpage}
\usepackage{lipsum}
\usepackage{setspace}

\usepackage{titletoc}

\usepackage[utf8]{inputenc}
\usepackage[english]{babel}
\usepackage{csquotes}
\usepackage[
sorting=nyc,
backend=biber, 
style=authoryear, 
uniquelist=false,
natbib,
block=ragged,
maxcitenames=2,
maxbibnames=99]{biblatex}

\DeclareNameFormat{always-init}{%
	\ifnumequal{\value{uniquename}}{2}
	{\usebibmacro{name:family-given}
		{\namepartfamily}
		{\namepartgiven}
		{\namepartprefix}
		{\namepartsuffix}}
	{\usebibmacro{name:family-given}
		{\namepartfamily}
		{\namepartgiveni}
		{\namepartprefix}
		{\namepartsuffix}}%
	\usebibmacro{name:andothers}}

\DeclareNameAlias{author}{always-init}
\DeclareNameAlias{editor}{always-init}

\DeclareSortingNamekeyTemplate{
	\keypart{\namepart{family}}
	\keypart{\namepart{prefix}}
	\keypart{\namepart{given}}
	\keypart{\namepart{suffix}}}

\DeclareSortingTemplate{nyc}{
	\sort{
		\field{presort}
	}
	\sort[final]{
		\field{sortkey}
	}
	\sort{
		\field{sortname}
		\field{author}
		\field{editor}
		\field{translator}
		\field{sorttitle}
		\field{title}
	}
	\sort{
		\field{sortyear}
		\field{year}
	}
	\sort{\citeorder}
}

\AtEveryBibitem{\clearfield{month}}
\AtEveryBibitem{\clearfield{day}}
\AtEveryBibitem{\clearfield{url}}
\AtEveryBibitem{\clearfield{urlyear}}
\AtEveryBibitem{\clearfield{note}}
\AtEveryBibitem{\clearfield{language}}

\DeclareNameWrapperFormat{labelname:poss}{#1's}

\newrobustcmd*{\posscitealias}{%
	\AtNextCite{%
		\DeclareNameWrapperAlias{labelname}{labelname:poss}}}

\newrobustcmd*{\posscite}{%
	\posscitealias
	\textcite}

\newrobustcmd*{\Posscite}{\bibsentence\posscite}

\newrobustcmd*{\posscites}{%
	\posscitealias
	\textcites}

\usepackage{ragged2e} 
\usepackage{titling} 
\usepackage{titlesec} 
\usepackage{amsthm, amsmath, amssymb} 
\usepackage{mathrsfs} 
\usepackage{float} 
\usepackage{graphicx, multicol} 
\usepackage{xcolor} 
\usepackage{framed}
\usepackage[normalem]{ulem} 
\usepackage{fancyhdr} 
\usepackage{sectsty} 

\usepackage[toc, title]{appendix} 

\usepackage[labelfont=bf]{caption} 

\usepackage{amsfonts}
\usepackage{enumitem}
\usepackage{mathtools}
\usepackage{multicol}
\usepackage{color,soul}

\usepackage{makecell,tabularx}

\usepackage{array}
\newcolumntype{M}[1]{>{\centering\arraybackslash}m{#1}}
\usepackage{rotating}
\usepackage{setspace}

\usepackage[boxed]{algorithm2e}
\DontPrintSemicolon

\usepackage{lipsum}

\theoremstyle{definition}

\usepackage{tikz}

\newcounter{cases}
\newcounter{subcases}[cases]

\newcommand{\myfig}[4]{\begin{figure}[H]\centering \begin{center} \includegraphics[width=#1\textwidth]{#2} \caption{#3} \label{#4} \end{center} \end{figure}}

\newlength{\depthofsumsign}
\newcommand*\mystrut[1]{\vrule width0pt height0pt depth#1\relax}

\usepackage{xr}

\makeatletter
\newcommand*{\addFileDependency}[1]{
	\typeout{(#1)}
	\@addtofilelist{#1}
	\IfFileExists{#1}{}{\typeout{No file #1.}}
}\makeatother

\usepackage[final, colorlinks = true, 
linkcolor = black, 
citecolor = black,
filecolor = black,
breaklinks=true,
hypertexnames=false]{hyperref} 

\sectionfont{\centering} 
\subsectionfont{\centering} 
\subsubsectionfont{\centering\itshape} 

\makeatletter
\def\blankfootnote{\xdef\@thefnmark{}\@footnotetext}
\makeatother

\renewenvironment{abstract}
{\small
\begin{center}
	\bfseries \abstractname\vspace{-.5em}\vspace{0pt}
\end{center}
\list{}{
	\setlength{\leftmargin}{0cm}
	\setlength{\rightmargin}{\leftmargin}%
}%
\item\relax}
{\endlist}

\title{A stochastic field theory for the evolution of quantitative traits in finite populations}
\author{Ananda Shikhara Bhat$^{1,2,*}$}
\date{}
\begin{document}
\maketitle
\vspace{-5em}
\begin{singlespace}
\noindent{\footnotesize$^{1}$ Department of Biology, Indian Institute of Science Education and Research Pune, Maharashtra-411008, India; $^{2}$ Centre for Ecological Sciences, Indian Institute of Science, Bengaluru, Karnataka-560012, India}
\end{singlespace}

\begin{abstract}
{
\begin{singlespace}
Infinitely many distinct trait values may arise in populations bearing quantitative traits, and modelling their population dynamics is thus a formidable task. While classical models assume fixed or infinite population size, models in which the total population size fluctuates due to demographic noise in births and deaths can behave qualitatively differently from constant or infinite population models due to density-dependent dynamics. In this paper, I present a stochastic field theory for the eco-evolutionary dynamics of finite populations bearing one-dimensional quantitative traits. I derive stochastic field equations that describe the evolution of population densities, trait frequencies, and the mean value of any trait in the population. These equations recover well-known results such as the replicator-mutator equation, Price equation, and gradient dynamics in the infinite population limit. For finite populations, the equations describe the intricate interplay between natural selection, noise-induced selection, eco-evolutionary feedback, and neutral genetic drift in determining evolutionary trajectories. My work uses ideas from statistical physics, calculus of variations, and SPDEs, providing alternative methods that complement the measure-theoretic martingale approach that is more common in the literature.
\end{singlespace}
}
\end{abstract}
\vspace{0.5em}

\blankfootnote{$*$\textit{Current Affiliation:} Institute of Organismic and Molecular Evolution (iomE) and Institute for Quantitative and Computational Biosciences (IQCB), Johannes Gutenberg University, 55128 Mainz, Germany. E-mail for correspondence: \href{mailto:abhat@uni-mainz.de}{abhat@uni-mainz.de}}

\begin{singlespace}
\end{singlespace}
\vspace{2em}
\begin{singlespace}
\end{singlespace}

\clearpage
\newpage
\begin{refsection}
\section*{Introduction}
\addcontentsline{toc}{section}{Introduction}

The success of the Modern Synthesis~\citep{provine_origins_2001,walsh_darwins_2014} illustrates the value of  abstract mathematical modelling in evolutionary biology. Several major modelling paradigms of eco-evolutionary dynamics --- such as evolutionary game theory and adaptive dynamics --- as well as the standard equations of population genetics and quantitative genetics, can be recovered (in a very general sense) from a small number of `fundamental' equations such as the replicator-mutator equation and Price equation~\citep{page_unifying_2002,lion_theoretical_2018,lehtonen_price_2018}. Historically, these `fundamental equations' have been formulated in deterministic terms through difference equations and ordinary/partial differential equations~\citep{page_unifying_2002,lion_theoretical_2018,lehtonen_price_2018}, and stochastic effects due to finite population sizes have been studied through more specific models such as the Wright-Fisher or Moran process~\citep{ewens_mathematical_2004}. 

Most classic stochastic models in population and quantitative genetics, such as the Wright-Fisher and  Moran models, assume the total number of individuals in the population either remains strictly constant or varies deterministically according to `top-down' rules that allow us to define the notion of a constant `effective population size'~\citep{lambert_population_2010}. However, the total population size in real populations is often an emergent property of individual-level ecological and demographic processes~\citep{metcalf_why_2007,lambert_population_2010}, and changes in the total population size due to these processes can cause qualitative changes in evolutionary dynamics relative to the expectations of constant population size models due to density-dependence~\citep{lambert_population_2010,papkou_hostparasite_2016,kokko_can_2017,mazzolini_universality_2023,de_vries_extrinsic_2023}. For instance, one important consequence of stochastic fluctuations of total population size is `noise-induced selection', an evolutionary force that can reverse the direction of evolution predicted by natural selection~\citep{gillespie_natural_1974,parsons_consequences_2010,constable_demographic_2016,mcleod_social_2019,week_white_2021,kuosmanen_turnover_2022,mazzolini_universality_2023,bhat_eco-evolutionary_2024}. Finite population models in population biology also typically incorporate stochasticity by adding noise to a `deterministic skeleton' rather than deriving the complete stochastic dynamics from first principles~\citep{coulson_skeletons_2004,lambert_population_2010,doebeli_towards_2017}.  Unfortunately, incorporating stochasticity in this manner is known to sometimes yield inconsistent predictions that disappear if the stochastic dynamics of finite population systems are instead systematically derived from individual-based rules~\citep{black_stochastic_2012,strang_how_2019}.

Stochastic individual-based models, in which (probabilistic) rules are specified at the level of the individual and population level dynamics are systematically derived from first principles, provide a natural way to describe eco-evolutionary dynamics of finite populations from first principles and avoid the potential pitfalls of ad-hoc implementations of stochasticity on one hand, and those of ignoring density-dependent ecological/demographic processes on the other. The process of systematically deriving equations of eco-evolutionary population dynamics from demographic first principles also provides a more mechanistic description of evolutionary dynamics~\citep{lambert_population_2010,doebeli_towards_2017} because `all paths to fitness lead through demography'~\citep{metcalf_all_2007}. While such a mechanistic description of eco-evolutionary dynamics can be carried out for populations bearing discrete traits without using too much mathematical machinery~\citep{parsons_consequences_2010,kuosmanen_turnover_2022,bhat_eco-evolutionary_2024}, studies that work with quantitative traits are currently grounded in the theory of measure-valued branching processes and their characterization via martingale theory and related fields~\citep{fournier_microscopic_2004,champagnat_unifying_2006,champagnat_evolution_2007,champagnat_individual_2008,champagnat_polymorphic_2011, week_white_2021, boussange_eco-evolutionary_2022}. As such, working with quantitative trait stochastic individual-based models analytically currently requires a considerable mathematical background in stochastic analysis and measure theory.

It has long been recognized that the equations governing the population level behavior of Markov processes based on stochastic individual-based dynamics are often very similar to the equations describing the behavior of many interacting particles in statistical physics~\citep{gardiner_stochastic_2009,black_stochastic_2012} and quantum mechanics~\citep{martin_statistical_1973,hochberg_effective_1999,baez_quantum_2018}. As a consequence, powerful heuristic tools originally developed in physics can be leveraged, under many situations, to study the behavior of systems in which a large number of individuals interact in a stochastic manner~\citep{martin_statistical_1973,doi_second_1976,peliti_path_1985,hochberg_effective_1999,thomas_system_2014,chow_path_2015,weber_master_2017,baez_quantum_2018}. Indeed, various tools from statistical and quantum mechanics have already been successfully applied to study stochasticity in biological populations~\citep{odwyer_integrative_2009,de_vladar_contribution_2011,black_stochastic_2012,schraiber_path_2014}. However, most studies that apply ideas from statistical physics to biological populations focus on modelling finite-dimensional systems. In contrast, populations bearing quantitative traits must be characterized by a function or distribution, and the object describing the state of the system at any given point is thus, in general, infinite-dimensional. In infinite dimensions, the analogy between statistical/quantum mechanics and Markov processes becomes an analogy between statistical/quantum field theory and infinite-dimensional stochastic processes such as SPDEs~\citep{hochberg_effective_1999,garcia-ojalvo_noise_1999}. While such `stochastic field equations' have been introduced to theoretical biology in the context of neurobiology~\citep{buice_field-theoretic_2007,bressloff_stochastic_2010,coombes_neural_2014} and collective motion~\citep{o_laighleis_minimal_2018}, they are as yet largely unused in population biology. 

In this paper, I present a general approach to modelling the evolution of one-dimensional quantitative traits in an arbitrary closed finite, fluctuating population starting from the demographic first principles of birth and death. My approach consists of describing the population as a stochastic `field' (function over space of allowed traits and time), assuming that birth and death rates scale according to a population size measure~\citep{czuppon_understanding_2021}, and then using ideas from statistical physics to derive stochastic equations that describe how this field changes over time when the carrying capacity is not too small. My framework largely only uses tools from calculus, calculus of variations, and some heuristics for spacetime white noise. It also yields stochastic partial differential equations (SPDEs) that are more amenable to direct attack using tools from statistical and quantum field theory than the more `analytic' formulation in terms of martingale problems. As such, my work is intended to complement the rigorous framework presented in previous studies~\citep{champagnat_unifying_2006,champagnat_individual_2008} with an alternative, heuristic formalism. In the next section, I present the general fomalism and describe the exact stochastic population dynamics via a functional master equation (Kolmogorov forward equation). These dynamics can then be approximated via an infinite-dimensional system size approximation to arrive at SPDEs describing the stochastic dynamics of biologically important quantities such as population abundances, trait frequencies, and trait means. I present the main results and their biological implications in the main text, and delegate detailed derivations to the Supplementary.

\section*{General field equation formalism}
\addcontentsline{toc}{section}{General field equation formalism}

Consider a closed population of individuals bearing a trait that takes values in a set $\mathcal{T} \subseteq \mathbb{R}$. I assume that the trait value of an individual cannot change over its lifetime. I will say that individuals that have the same value of the trait are of the same `type' and are all identical for the purposes of our model. Each individual with a trait value $x \in \mathcal{T}$ can be characterized as a single Dirac delta mass centered at $x$, defined indirectly as the object which satisfies, for any one-dimensional real function $f$ and any set $A \subseteq \mathcal{T}$,
\begin{linenomath*}\begin{equation*}
		\int\limits_{A}f(y)\delta_{x}dy = 
		\begin{cases}
			f(x) & x \in A\\
			0 &  x \notin A 
		\end{cases}
\end{equation*}\end{linenomath*}
Intuitively, $\delta_x$ should be thought of as analogous to an indicator function: Given any set $A \subseteq \mathcal{T}$, integrating $\delta_x$ over $A$ returns 1 if $x \in A$ and 0 otherwise, and thus tells us whether or not the individual in question has a trait value that lies within the set $A$. Note that in physics notation, the Dirac mass centered at $x$ would be $\delta(y-x)$, where $y$ is a dummy variable, and the integral would be written $\int_A f(y) \delta(y-x) dy$.

If the population at any time $t$ consists of $N(t)$ individuals with trait values $\{x_1,x_2,\ldots,x_{N(t)}\}$, it can then be completely characterized (Fig. \ref{fig_infD_pop_description}) by the `stochastic field' (finite measure)
\begin{linenomath*}\begin{equation*}
		\nu_t = \sum\limits_{i=1}^{N(t)}\delta_{x_i}
\end{equation*}\end{linenomath*}
Thus, we are interested in formulating and studying a stochastic process taking values in
\begin{linenomath*}\begin{equation*}
		\mathcal{M}(\mathcal{T})= \left\{\sum\limits_{i=1}^{n}\delta_{x_i} \ | \ n \in \mathbb{N}, x_i \in \mathcal{T}\right\}
\end{equation*}\end{linenomath*}
The elements of $\mathcal{M}(\mathcal{T})$ are formally so-called `finite measures'. For our purposes, $\nu \in \mathcal{M}(\mathcal{T})$ can be thought of as analogous to a `density function' in the sense that for any subset $A \subset \mathcal{T}$ of the trait space, the quantity $\int_A\nu dx$ gives the number of individuals that have trait values that lie within the set $A$ in the population $\nu$. I will use the notation $\nu(x,t)$ to denote the field at time $t$ to help the reader remember this interpretation of $\nu$ as a `density' over the trait space. It is important to note that just like with probability densities, the `value of the field at trait value $x$' is undefined, and thus it only makes sense to speak of quantities of the form $\int f(x,\nu)\nu(x,t)dx$ for some function $f(x,\nu)$ (Note that in measure-theoretic notation, $\nu$ is viewed as a measure and thus integration of $f$ against $\nu$ is usually written $\langle f, \nu \rangle =  \int f(x,\nu)\nu(dx)$). In this sense, the variable $x$ in $\nu(x,t)$ is a dummy variable that will be integrated over, but I retain it in the interest of suggestive notation.  It is easy to see from the definition of the Dirac mass that when the population is given by $\nu(y,t) = \sum_{i=1}^{N(t)}\delta_{x_i}$, we have $\int_{\mathcal{T}}f(y)\nu(y,t)dy = \sum_{i=1}^{N(t)}f(x_i)$. Thus,  $\int_{\mathcal{T}}f(y)\nu(y,t)dy $ simply evaluates the function $f$ for each trait value $x_i$ that is currently present in the population.

\myfig{1}{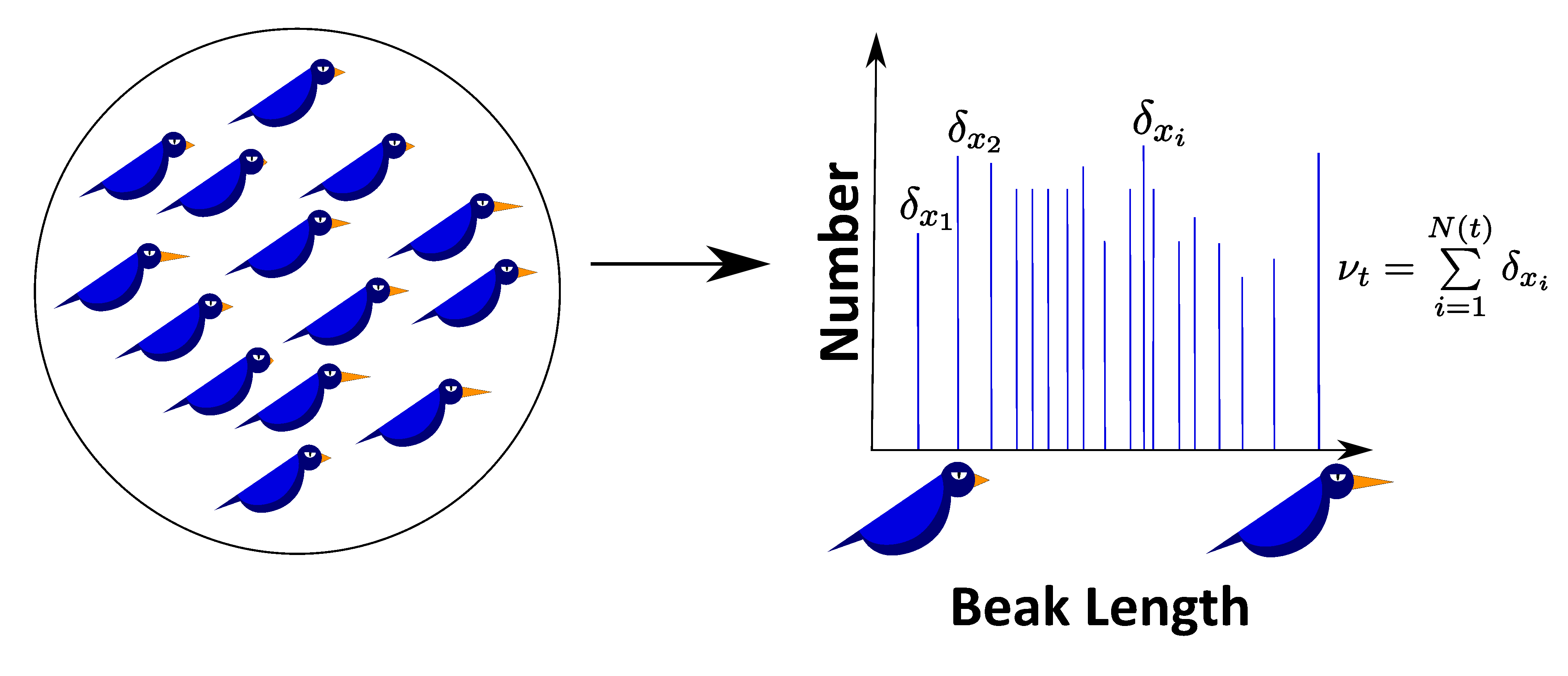}{\textbf{Schematic description of the state space of our stochastic process.} Consider a population of birds in which individuals have varying beak lengths. Since each individual has a single beak length, it can be characterized as a Dirac mass centered at its beak length. Thus, if the population has $N(t)$ individuals, it can be described as a distribution obtained as the sum of $N(t)$ (not necessarily distinct) Dirac masses.}{fig_infD_pop_description}
Now that we have described the population, we must define the rules for how it changes over time. I assume that the probability of observing two or more simultaneous events (births or deaths) at the same instant of time is negligible. Thus, the population changes in units of a single individual. Since we assumed the population is closed, changes in the population can now be described using two non-negative functionals $b(x|\nu)$ and $d(x|\nu)$ from $\mathcal{T} \times \mathcal{M}(\mathcal{T})$ to $[0,\infty)$ that specify the absolute (rather than per-capita) rate at which individuals with trait value $x$ are born and die respectively in a population $\nu$. That is, if we know that the population was in the state $\nu$ and we know that \emph{either a birth or a death} has occurred and changed the population to some state other than $\nu$, then the probability that the event which occurred is the birth of an individual whose phenotype is within the set $A \subset \mathcal{T}$ is given by
\begin{linenomath*}\begin{equation*}
		\mathbb{P}\big[\textrm{ Birth with offspring in $A$ } \big{|} \textrm{ something happened }\big] = \frac{1}{\mathcal{N}(\nu)}\int\limits_{A}b(x|\nu)dx
\end{equation*}\end{linenomath*}
and the probability that the event is the death of an individual whose phenotype is within the set $A$ is
\begin{linenomath*}\begin{equation*}
		\mathbb{P}\big[\textrm{ Death of an individual in $A$ } \big{|} \textrm{ something happened }\big] = \frac{1}{\mathcal{N}(\nu)}\int\limits_{A}d(x|\nu)dx
\end{equation*}\end{linenomath*}
where $\mathcal{N}(\nu) = \int_{\mathcal{T}}b(x|\nu)+d(x|\nu)dx$ is a normalizing constant that could depend on the current state $\nu$. I assume $\mathcal{N}(\nu)$ is always finite and non-zero. Our description of population dynamics is thus a `function'-valued Markov process (and more precisely a measure-valued birth-death process) with transition probabilities given by the birth and death rate functionals $b(x|\nu)$ and $d(x|\nu)$. I further assume that the total birth and death rates do not scale faster than linearly with the total population size, \emph{i.e.} that for a given population $\nu$, the quantities $\int_{\mathcal{T}}b(x|\nu)dx$ and $\int_{\mathcal{T}}d(x|\nu)dx$ are $\mathcal{O}(\int_{\mathcal{T}} \nu dx)$. This latter assumption will be used in the section below when conducting a system-size expansion.

Let us now define, for each $x \in \mathcal{T}$, two \emph{step operators} $\mathcal{E}_{x}^{\pm}$ defined by their action on any function $f(y,\nu):\mathcal{T}\times\mathcal{M}(\mathcal{T}) \to \mathbb{R}$ as:
\begin{linenomath*}\begin{equation*}
		\mathcal{E}_{x}^{\pm}f(y,\nu) =  f(y,\nu \pm \delta_x)
\end{equation*}\end{linenomath*}
In other words, the step operators $\mathcal{E}_{x}^{\pm}$ simply describe the effect of adding or removing a single individual with trait value $x$ from the population $\nu$.

Let $P(\nu,t | \nu_0, 0)$ be the (conditional) probability density function of the process, \emph{i.e.} the probability that the population is described by $\nu$ at time $t$ given that it is described by $\nu_0$ at time 0. From now on, I omit the conditioning and simply write $P(\nu,t)$ for notational simplicity. Recall that any change to the population must be through the birth or death of a single individual, \emph{i.e.} through addition or subtraction of a single Dirac mass. For any state $\nu \in \mathcal{M}(\mathcal{T})$, a population could end up in the state $\nu$ either through a birth of an individual $\delta_x$ in a population $\nu - \delta_x$, or through the death of an individual $\delta_x$ in a population $\nu + \delta_x$, for any possible $x \in \mathcal{T}$. Transitions of the form $\nu-\delta_x \to \nu$ occur through the birth of an individual with trait value $x$, and by the definition of our birth rates above, the transition rate from $\nu - \delta_{x}$ to $\nu$ is thus given by $b (x | \nu - \delta_x)$. To find the total transition probability into the state $\nu$, we must now `sum over' (integrate) the contributions of transitions due to births of individuals of all possible trait values $x \in \mathcal{T}$. Thus, the transition rate into the state $\nu$ at time $t$ due to births is
\begin{linenomath*}\begin{equation}
		\label{rate_in_births}
		R^{\textrm{births}}_{\textrm{in}}(\nu, t) = \underbrace{ \int\limits_{\mathcal{T}} }_{\substack{\text{`sum over'}\\\text{all possible $x$}}} \underbrace{b (x | \nu - \delta_x) }_{\substack{\text{Rate of}\\ {(\nu-\delta_x) \to \nu} \\ \text{transition} }} \underbrace{P(\nu - \delta_x, t)}_{\substack{\text{Probability of}\\\text{finding the} \\ \text{population $\nu-\delta_x$}}} \ dx = \int\limits_\mathcal{T}[\mathcal{E}^{-}_{x}b(x|\nu)P(\nu,t)]dx
\end{equation}\end{linenomath*}
where we have rewritten the term within the integral on the RHS of Eq. \ref{rate_in_births} as $b (x | \nu - \delta_x)P(\nu - \delta_x, t) = \mathcal{E}^{-}_{x}b(x|\nu)P(\nu,t)$ using the definition of the step operator $\mathcal{E}^{-}_{x}$. Similarly, the transition from $\nu+\delta_{x}$ to $\nu$ is through death of type $x$ individuals and thus has transition rate $d (x | \nu + \delta_x)$. The total transition probability into the population $\nu$ at time $t$ due to deaths of individuals can thus be written
\begin{linenomath*}\begin{equation}
		\label{rate_in_deaths}
		R^{\textrm{deaths}}_{\textrm{in}}(\nu, t) = \underbrace{ \int\limits_{\mathcal{T}} }_{\substack{\text{`sum over'}\\\text{all possible $x$}}} \underbrace{d (x | \nu + \delta_x) }_{\substack{\text{Rate of}\\ {(\nu+\delta_x) \to \nu} \\ \text{transition} }} \underbrace{P(\nu + \delta_x, t)}_{\substack{\text{Probability of}\\\text{finding the} \\ \text{population $\nu+\delta_x$}}} \ dx = \int\limits_\mathcal{T}[\mathcal{E}^{+}_{x}d(x|\nu)P(\nu,t)]dx
\end{equation}\end{linenomath*}
The transition rate out of $\nu$ to a state $\nu+\delta_x$ due to births of type $x$ individuals is $b(x|\nu)$, and transition out to a state $\nu - \delta_x$ due to death of type $x$ individuals is $d(x|\nu)$. Thus, the transition probabilities of exiting the state $\nu$ at time $t$ are given by
\begin{linenomath*}\begin{align}
		\label{rate_out_births}
		R^{\textrm{births}}_{\textrm{out}}(\nu, t) &=  \int\limits_\mathcal{T}b(x|\nu)P(\nu,t)dx\\
		\label{rate_out_deaths}
		R^{\textrm{births}}_{\textrm{out}}(\nu, t) &=  \int\limits_\mathcal{T}d(x|\nu)P(\nu,t)dx
\end{align}\end{linenomath*}
Now, the total probability flux through the state $\nu$ must be given by the difference between the rate of inflow and the rate of outflow of probability, or, in equations,
\begin{linenomath*}
	\begin{equation}
		\label{prob_flux_as_rate_diff}
		\underbrace{\frac{\partial P}{\partial t}(\nu,t)}_{\substack{\text{Rate of change}\\\text{of probability of}\\\text{the population being}\\\text{described by $\nu$}}} = \underbrace{\vphantom{\frac{\partial P}{\partial t}}\left[R^{\textrm{births}}_{\textrm{in}}(\nu, t) + R^{\textrm{deaths}}_{\textrm{in}}(\nu, t)\right]}_{\substack{\text{Total rate of `inflow'}\\\text{into the population $\nu$}\\ \text{due to births and deaths}}} - \underbrace{\vphantom{\frac{\partial P}{\partial t}}\left[R^{\textrm{births}}_{\textrm{out}}(\nu, t) + R^{\textrm{deaths}}_{\textrm{out}}(\nu, t)\right]}_{\substack{\text{Total rate of `outflow'}\\\text{from the population $\nu$}\\ \text{due to births and deaths}}}
	\end{equation}
\end{linenomath*} 
Substituting Eqs \ref{rate_in_births} - \ref{rate_out_deaths} into Eq. \ref{prob_flux_as_rate_diff} and rearranging, we thus see that $P(\nu,t)$ must satisfy:
\begin{linenomath*}\begin{equation}
		\label{maintext_unnormalized_M_equation}
		\frac{\partial P}{\partial t}(\nu,t) = \int\limits_{\mathcal{T}}\left[(\mathcal{E}^{-}_{x}-1)b(x|\nu)P(\nu,t) + (\mathcal{E}^{+}_{x}-1)d(x|\nu)P(\nu,t)\right]dx
\end{equation}\end{linenomath*}
Equation \ref{maintext_unnormalized_M_equation} completely describes the stochastic evolution of the population, and can be thought of as an infinite-dimensional `master equation'~(\cite{van_kampen_stochastic_1981}, Equation 5.1.5) or Kolmogorov Forward Equation~(\cite{karatzas_brownian_1998}, Equation 5.1.6).

\subsection*{The population density field}
\addcontentsline{toc}{subsection}{The population density field}

On ecological grounds, I assume that that the birth-death process admits a carrying capacity, or more generally a \emph{population size measure}~\citep{czuppon_understanding_2021} $K > 0$ such that the expected population growth rate of every type is negative whenever the total population size exceeds $K$. In other words, I assume that the functionals $b(x|\nu)$ and $d(x|\nu)$ are such that there exists a $K > 0$ such that for any set $A \subseteq \mathcal{T}$, $\int_A \left[ \ b(x|\nu) - d(x|\nu) \ \right] dx < 0$ whenever $\int_{\mathcal{T}}\nu dx > K$. In this case, we expect the stochastic process to remain in the domain where the total population size $\int_{\mathcal{T}}\nu dx$ is $\mathcal{O}(K)$. Thus, $K = \infty$ corresponds to an infinitely large population. By dividing $\nu(\cdot,t)$ by $K$, we can now obtain a new field $\phi(x,t)=\nu(x,t)/K$. For any set $A \subset \mathcal{T}$, $\int_A\phi(x,t)dx$ gives the `population density' of individuals that have trait values that lie within the set $A$. I will call $\phi$ the population density field. The field at time $t$ is defined as
\begin{linenomath*}\begin{equation*}
		\phi(\cdot,t) \coloneqq \frac{1}{K}\nu(\cdot,t) = \frac{1}{K}\sum\limits_{i=1}^{N(t)}\delta_{x_i}
\end{equation*}\end{linenomath*}
Note that since the total population size $\int_{\mathcal{T}}\nu(x,t) dx$ is $\mathcal{O}(K)$, the total population density $\int_{\mathcal{T}}\phi(x,t) dx$ is $\mathcal{O}(1)$. The parameter $K$, along with the scaling assumptions on the birth and death rate functionals, are such that, for any $t > 0$, $\int_{\mathcal{T}} \phi(x,t) dx \to 1$ as $K \to \infty$. Thus, the limit $K \to \infty$ corresponds to finite population densities but infinite population sizes. Mathematically, I assume we can find two non-negative $\mathcal{O}(1)$ functionals $b_K$ and $d_K$ such that the birth and death rate functionals $b(x|\nu)$ and $d(x|\nu)$ can be rewritten as:
\begin{linenomath*}\begin{equation}
		\label{pop_density_BD_variable_transform}
		\begin{aligned}
			b(x|\nu) &= Kb_K(x|\nu/K) = Kb_K(x|\phi)\\
			d(x|\nu) &= Kd_K(x|\nu/K) = Kd_K(x|\phi)
		\end{aligned}
\end{equation}\end{linenomath*}
. The new stochastic field $\{\phi(\cdot,t)\}_{t\geq0}$ takes values in 
\begin{linenomath*}\begin{equation*}
		\mathcal{M}_{K}(\mathcal{T}) \coloneqq \left\{\frac{1}{K}\sum\limits_{i=1}^{n}\delta_{x_i} \ | \ n \in \mathbb{N}, x_i \in \mathcal{T}\right\}
\end{equation*}\end{linenomath*}
As before, $\phi(x,t)$ can be thought of as analogous to a `probability density function' for the population density in the sense that the population density of individuals with trait values that lie within the (infinitesimal) interval $(x, x + dx)$ is informally $\phi(x,t)dx$. If $K$ is large, each individual contributes a negligible amount to the total density field and the field as a whole begins to look approximately continuous (over the trait space), allowing us to now speak about the value of the field at a particular trait value $x$ instead of merely speaking about the density of individuals within subsets of the trait space. I take this continuity as an assumption, but it can be shown to rigorously hold under various technical conditions using more sophisticated mathematical tools~\citep{champagnat_unifying_2006,week_white_2021}. Let $P(\phi,t | \phi_0, 0)$ be the probability that the population density field is given by $\phi \in \mathcal{M}_K(\mathcal{T})$ at time $t$ if the stochastic process was initialized with the field $\phi_0 \in \mathcal{M}_K(\mathcal{T})$ at time $0$. As before, I omit the conditioning below for notational simplicity. We are interested in finding an equation for how $P(\phi,t)$ changes over time.

\subsection*{Functional forms of the birth and death rates}
\addcontentsline{toc}{subsection}{Functional forms of the birth and death rates}

I assume that the birth and death functions take the form:
\begin{linenomath*}\begin{equation}
		\label{BD_defns}
		\begin{aligned}
			b_K(x|\phi) &=  \phi(x,t)b^{\textrm{(ind)}}(x|\phi) + \mu Q(x|\phi)\\
			d_K(x|\phi) &= \phi(x,t)d^{\textrm{(ind)}}(x|\phi)
		\end{aligned}
\end{equation}\end{linenomath*}
where $\mu \geq 0$ is  a constant and $b^{\textrm{(ind)}}(x|\phi)$, $d^{\textrm{(ind)}}(x|\phi)$, and $Q(x|\phi)$ are all $\mathcal{O}(1)$ and continuous in both $x$ and $\phi$. Here, the functionals $b^{\textrm{(ind)}}(x|\phi)$ and $d^{\textrm{(ind)}}(x|\phi)$ can be thought of as describing the birth and death rate of type $x$ organisms in a population $\phi$ at an individual (per-capita) level. The term $Q(x|\phi)$ describes contributions to the birth rate that are not of the form $\phi F[x|\phi]$. If at some time $t$ there are no type $x$ individuals in the population ($\phi(x,t) = 0$), type $x$ individuals may still arise in the population due to immigration or mutations of other types during birth. However, such births cannot be incorporated into $b^{\textrm{(ind)}}$ since the product $\phi(x,t)b^{\textrm{(ind)}}(x|\phi)$ vanishes whenever $\phi(x,t) = 0$. For instance, if individuals of type $x$ are born due to mutations at birth of a different type $y$, the contribution to the birth rate of type $x$ depends on a mutation rate and on the value of the density field at $y$ (i.e. on $\phi(y,t)$), but does \emph{not} depend on the density of type $x$ individuals (i.e. $\phi(x,t)$). For simplicity, I will henceforth assume that the population is closed and $Q$ describes the effects of potential mutational effects during birth, with $\mu \geq 0$ being a constant mutation rate. Note that we do not need to include such a term for the death rate, since we must necessarily have $d_K(x|\phi) = 0$ when $\phi(x,\cdot) = 0$ to avoid negative population density values and thus mutation/emigration that leads to loss of individuals can be subsumed into $d^{\textrm{(ind)}}$. The functionals $b^{\textrm{(ind)}}(x|\phi)$, $d^{\textrm{(ind)}}(x|\phi)$, and $Q(x|\phi)$ may be quite complex (as long as they are all bounded for all $x, \phi$ as $t \to \infty$) and could in principle model several ecological phenomena.

I define $w(x|\phi)$, the \emph{Malthusian fitness} of type $x$ in a population $\phi$ as
\begin{linenomath*}\begin{equation}
		w(x|\phi) \coloneqq b^{\textrm{(ind)}}(x|\phi) - d^{\textrm{(ind)}}(x|\phi)
\end{equation}\end{linenomath*}
In words, $w(x|\phi)$ is a measure of the (stochastic) growth rate of type $x$ individuals in the population defined by the population density field $\phi$ due to non-mutational effects. I also define $\tau(x|\phi)$, the \emph{per-capita turnover} rate of type $x$ in a population $\phi$, as
\begin{linenomath*}\begin{equation}
		\tau(x|\phi) \coloneqq b^{\textrm{(ind)}}(x|\phi) + d^{\textrm{(ind)}}(x|\phi)
\end{equation}\end{linenomath*}
The quantity $\tau(x|\phi)$ is a measure of the expected total number of (stochastic) changes to the density field at point $x$ in a population $\phi$ due to non-mutational effects.

\subsection*{Statistical measures for type-level quantities}
\addcontentsline{toc}{subsection}{Statistical measures for type-level quantities}

So far, we have been speaking entirely in terms of population densities. However, evolution is not in terms of population densities, but in terms of trait frequencies. To track population sizes, I compute the scaled population size $N_K$ as
\begin{linenomath*}\begin{equation}
		\label{pop_size_defn}
		N_K(t) \coloneqq \int\limits_{\mathcal{T}}\phi(x,t)dx = \frac{1}{K} \int\limits_{\mathcal{T}}\nu(x,t)dx
\end{equation}\end{linenomath*}
Thus, $KN_K(t)$ is the total population size at time $t$. When the population is at carrying capacity, $N_K = 1$. Further, recall that we assumed $\int_{\mathcal{T}} \phi(x,t) dx = N_K(t) \to 1$ as $K \to \infty$ for any fixed $t$. 

I now define the \emph{trait frequency field} $p(x,t)$, a stochastic field given by
\begin{linenomath*}\begin{equation}
		\label{freq_defn}
		p(x,t) \coloneqq \frac{\nu(x,t)}{\int\limits_{\mathcal{T}}\nu(y,t)dy} =  \frac{\phi(x,t)}{N_K(t)}
\end{equation}\end{linenomath*}
Integrating the $p(x,t)$ field in the $x$ variable over any set $A \subseteq \mathcal{T}$ gives us the frequency of individuals bearing trait values that lie in the set $A$.

Now, let $f(x|\phi):\mathcal{T} \times \mathcal{M}(\mathcal{T}) \to \mathbb{R}$ be a real function. For example, $f$ could describe a phenotype, a quantity such as fitness or turnover rate, or simply a label defined at each trait value $x$. Given any such type-level quantity, We can define the mean value of $f$ in the population $\phi$ at time $t$ as
\begin{linenomath*}\begin{equation}
		\label{defn_mean_value}
		\overline{f}(t) = \int\limits_{\mathcal{T}}f(x|\phi)p(x,t)dx
\end{equation}\end{linenomath*}
the statistical covariance of two quantities $f$ and $g$ as
\begin{linenomath*}\begin{equation}
		\label{defn_covariance}
		\textrm{Cov}(f,g) = \overline{fg} - \overline{f}\overline{g}
\end{equation}\end{linenomath*}
and the statistical variance of a quantity $f$ as $\sigma^2_f = \textrm{Cov}(f,f)$. Note that these three quantities are all statistical measures that describe how traits are distributed in a given population $\phi$. They are distinct from the \emph{probabilistic} expectation, variance, and covariance obtained by integrating over an ensemble of populations that represent different realizations of the stochastic process. I denote the probabilistic expectation by $\mathbb{E}[\cdot]$.
\begin{center}
	\begin{table}
		\centering
		\footnotesize
		\begin{tabular}{  M{0.16\textwidth} | M{0.85\textwidth} }
			\hline
			\textbf{Symbol} & \textbf{Meaning} \\
			\hline
			$\mathcal{T}$ & Trait space, assumed a subset of $\mathbb{R}$.\\
			$\delta_x$ & Dirac mass centered at $x \in \mathcal{T}$. We will use this to characterize a single individual.\\
			$\nu(y,t)$ & A stochastic field describing the population at time $t$. If the population at time $t$ consists of individuals having trait values $\{x_1, x_2, x_3, \ldots\}$, then $\nu(y,t) = \sum_i \delta_{x_i}$ describes the entire population. \\
			$\mathcal{M}(\mathcal{T})$ & The set $\left\{\sum\limits_{i=1}^{n}\delta_{x_i} \ | \ n \in \mathbb{N}, x_i \in \mathcal{T}\right\}$. This is the state space of our stochastic process.\\
			$b(x|\nu), d(x|\nu)$ & Birth and death rate functionals for the birth-death process $\{\nu(\cdot,t)\}_{t \geq 0}$.\\
			$K$ & Population size measure~\citep{czuppon_understanding_2021}. A non-negative number that controls the expected total population size. $K \to \infty$ yields the infinite population size (but finite population density) limit.\\
			$\phi(y,t)$ & Population density field, $\nu(y,t)/K$\\
			$\mathcal{M}_K(\mathcal{T})$ & The set $\left\{\frac{1}{K}\sum\limits_{i=1}^{n}\delta_{x_i} \ | \ n \in \mathbb{N}, x_i \in \mathcal{T}\right\}$. This is the state space of our rescaled stochastic process $\{\phi(\cdot,t)\}_{t \geq 0}$.\\
			$N_K(t)$ & The rescaled population size $\int_{\mathcal{T}}\phi(y,t)dy$. $KN_K(t)$ is the total population size at time $t$.\\
			$p(x,t)$ & The trait frequency field $\phi(x,t)/N_K(t)$.\\
			$b_K(x|\phi), d_K(x|\phi)$ & Birth and death rate functionals for the rescaled process $\{\phi(\cdot,t)\}_{t \geq 0}$. I assume the functional forms $b_K(x|\phi) =  \phi(x,t)b^{\textrm{(ind)}}(x|\phi) + \mu Q(x|\phi), d_K(x|\phi) = \phi(x,t)d^{\textrm{(ind)}}(x|\phi)$\\
			$P(\nu, t)$ & Shorthand for $P(\nu, t | \nu_0, 0)$, Probability of finding the population in a state $\nu$ at time $t$ if it begins in a state $\nu_0$ at time $0$.\\
			$b^{\textrm{(ind)}}, d^{\textrm{(ind)}}$ & Per-capita birth and death rates, excluding potential mutational effects (see below)\\
			$Q(x|\phi)$ & Function to model contributions to birth rate that cannot be written in terms of per-capita rates. The strength is parameterized by a constant $\mu > 0$. The analogy is with mutation (parametrized by a mutation rate) or immigration (parameterized by a migration rate). See main text for details. \\
			$w(x|\phi)$ & Malthusian fitness $b^{\textrm{(ind)}}(x|\phi) - d^{\textrm{(ind)}}(x|\phi)$ of trait value $x$ in a population $\phi$.\\
			$\tau(x|\phi)$ & Per-capita turnover rate $b^{\textrm{(ind)}}(x|\phi) + d^{\textrm{(ind)}}(x|\phi)$ of trait value $x$ in a population $\phi$.\\
			$\overline{f}$ & Statistical mean of $f(x|\phi)$ in the population, computed as $\int_{\mathcal{T}}f(x|\phi)p(x,t)dx$.\\
			$\textrm{Cov}(f,g)$ & Statistical covariance between $f(x|\phi)$ and $g(x|\phi)$ in the population, computed as $\overline{fg} - \overline{f}\overline{g}$.\\
			$\dot{W}(x,t)$ & A spacetime white noise process on $\mathcal{T} \times [0,\infty)$ .\\
			\hline
		\end{tabular}
		
		\caption{Table of Notation}
	\end{table}
\end{center}

\section*{Results}
\addcontentsline{toc}{section}{Results}

In principle, Eq.~\ref{maintext_unnormalized_M_equation} exactly describes the complete stochastic population dynamics of our population of interest. However, in practice, the exact stochastic process is usually much too complicated to be studied directly on an exact level. Instead, we will look for approximate continuous field equations that describe the density field $\phi(x,t)$, the trait frequency field $p(x,t)$, and the mean value $\overline{f}$ of any field describing a type-level quantity $f(x|\phi)$ (of particular interest will be $\overline{x}$, the mean value of the trait itself). I relegate the detailed calculations to the supplementary material, which is entirely mathematically self-contained (except for using Eq.~\ref{maintext_unnormalized_M_equation} as a starting point). I provide the big picture (on a mathematical level) outlining the reasoning behind the calculations in supplementary section~\ref{App_outline}. In the main text, I instead focus on discussing the major results, their biological implications and interpretations, and connections to various existing formal descriptions of the evolution of quantitative traits.

\subsection*{A field equation for population densities}
\addcontentsline{toc}{subsection}{A field equation for population densities}

In supplementary section \ref{App_system_size}, I obtain an approximate equation describing the behaviour of the density field $\phi$ using an infinite-dimensional analog of the system-size expansion~(\cite{gardiner_stochastic_2009}, Chapter 13), also called the diffusion approximation in the population genetics literature~\citep{crow_introduction_1970,ewens_mathematical_2004}. I show that if $K$ is reasonably large, $P(\phi,t)$ approximately evolves according to the equation:
\begin{linenomath*}\begin{equation}\label{density_FPE}
		\resizebox{\textwidth}{!}{$\displaystyle
			\frac{\partial P}{\partial t}(\phi,t) = \int\limits_{\mathcal{T}}\left[-
			\frac{\delta}{\delta\phi(x)} \big{\{} \big{[}\phi(x) w(x|\phi) + \mu Q(x|\phi)\big{]} P(\phi,t) \big{\}} + \frac{1}{2K}\frac{\delta^2}{\delta\phi(x)^2}\big{\{}\big{[}\phi(x) \tau(x|\phi) + \mu Q(x|\phi)\big{]}P(\phi,t)\big{\}}\right]dx$}
\end{equation}\end{linenomath*}
where I have suppressed the $t$ dependence of $\phi$ for conciseness. Here, $\delta F/ \delta \rho$ denotes the \emph{functional derivative} of the functional $F$ with respect to the function $\rho$, defined indirectly as the unique object that satisfies for any function $\xi$
\begin{linenomath*}\begin{equation}
		\label{functional_derivative_defn}
		\int\frac{\delta F}{\delta \rho(x)}\xi(x)dx = \lim_{h \to 0} \frac{F[\rho + h\xi]-F[\rho]}{h}.
\end{equation}\end{linenomath*}
Equation \ref{density_FPE} is a functional Fokker-Planck equation~(\cite{gardiner_stochastic_2009}, Equation 13.1.25; \cite{garcia-ojalvo_noise_1999}, Equation 2.54) or Kolmogorov forward equation~(\cite{karatzas_brownian_1998}, Equation 5.1.6) for $P(\phi,t)$ , the probability of finding the population in a state $\phi$ at time $t$. We now recall that a finite-dimensional Markov process whose density is described by a Fokker-Planck equation can always equivalently be represented as the solution to an It\^o stochastic differential equation~(\cite{gardiner_stochastic_2009}, section 4.3.5). Exactly analogously, infinite-dimensional Markov processes whose density functions are described by functional Fokker-Planck equations can be represented as solutions to It\^o stochastic partial differential equations~(\cite{konno_stochastic_1988}, theorem 1.4; \cite{dawson_stochastic_2000}, theorem 1.2 with $g \equiv 0, \gamma = \sigma^2 = 1, \epsilon^2 = 2$). Thus, the stochastic process whose probability density is described by Eq. \ref{density_FPE} must satisfy the stochastic partial differential equation (SPDE):
\begin{linenomath*}\begin{equation}
		\label{density_SPDE}
		\frac{\partial \phi}{\partial t}(x,t) = \left[\phi(x,t) w(x|\phi) + \mu Q(x|\phi)\right] + \frac{1}{\sqrt{K}}\sqrt{\phi(x,t) \tau(x|\phi) + \mu Q(x|\phi)}\dot{W}(x,t)
\end{equation}\end{linenomath*}
where $\dot{W}(x,t)$ is the \emph{spacetime white noise} on $\mathcal{T}\times [0,\infty)$, defined indirectly~\citep{pardoux_stochastic_2021} as the object that satisfies for any two square-integrable functions $f, g$ on $\mathcal
{T} \times [0,\infty)$ and any time $t > 0$:
\begin{linenomath*}\begin{equation}
		\begin{aligned}
			\mathbb{E}\left[\int\limits_{0}^{\hphantom{t}t}\int\limits_{\mathcal{T}}f(u,s)\dot{W}(u,s)duds\right] &= 0\\
			\mathbb{E}\left[\int\limits_{0}^{\hphantom{t}t}\int\limits_{\mathcal{T}}f(u,s)\dot{W}(u,s)duds\int\limits_{0}^{\hphantom{t}t}\int\limits_{\mathcal{T}}g(u,s)\dot{W}(u,s)duds\right] &= \int\limits_{0}^{\hphantom{t}t}\int\limits_{\mathcal{T}}f(u,s)g(u,s)duds.
		\end{aligned}
\end{equation}\end{linenomath*} 
\cite{week_white_2021} provide an excellent introduction to spacetime white noise processes. In supplementary section \ref{sec_Fourier}, I show how the equation for the density field can be combined with a weak noise approximation and Fourier techniques to study phenotypic clustering and adaptive diversification/speciation. The analytic pipeline for studying phenotypic clustering via Fourier techniques has already been developed for some specific models in previous studies~\citep{rogers_demographic_2012,rogers_modes_2015}, but to the best of my knowledge, supplementary section \ref{sec_Fourier} provides the first general treatment.

\subsubsection*{The infinite population limit}

Taking the infinite population limit ($K \to \infty$) in equation \ref{density_SPDE} yields a deterministic process whose evolution is described by the PDE:
\begin{linenomath*}\begin{equation}
		\label{deterministic_traj}
		\frac{\partial \phi}{\partial t}(x,t) = \underbrace{\phi(x,t)w(x|\phi)}_{\substack{\text{Growth rate due to}\\\text{ecological interactions}}}+\underbrace{\mu Q(x|\phi)}_{\substack{\text{Additional growth rate}\\\text{due to mutations}}}
\end{equation}\end{linenomath*}
Equation \ref{deterministic_traj} describes the change in population densities as the sum of two terms. The first term is the difference between the per-capita birth and death rates of type $x$ individuals ($w(x|\phi)$) multiplied by the current population density at the point $x$, and represents growth due to ecological interactions in the absence of mutation; This can be seen by comparing equation \ref{deterministic_traj} (with $\mu = 0$) to one-dimensional ecological models of the form $\dot{N}_t = N_tf(N_t)$ (ex: $f(N_t) = r$ gives exponential growth, $f(N_t) = (1-N_t/K)$ gives logistic growth, and so on). The second term on the RHS of Eq. \ref{deterministic_traj} describes the effects of mutation on growth rate; This term is always non-negative because we only incorporated the effects of mutation in the birth rate in Eq. \ref{BD_defns}. Models of this form are precisely the non-spatial `PDE models' discussed in studies of adaptive diversification~\citep{doebeli_adaptive_2011}. Equation \ref{deterministic_traj} is also the starting point of `oligomorphic dynamics' if one assumes the population is composed of a small number of `morphs', \emph{i.e.} $\phi(x,t) = \sum\limits_{k=1}^{S} n_{k}(t)\phi_k(x,t)$, where $n_{k} \geq 0$ is the abundance of the $k$\textsuperscript{th} morph, $\phi_k(x,t)$ is the phenotypic distribution of the $k$\textsuperscript{th} morph, and $S$ is the total number of distinct morphs in the population~\citep{sasaki_oligomorphic_2011,lion_extending_2023}. Finally, equations of the form Eq. \ref{deterministic_traj} have also recently been proposed as models to study the effects of intraspecific trait variation in ecological communities~\citep{nordbotten_dynamics_2020,wickman_theoretical_2023}.

\subsection*{A stochastic replicator equation for the trait frequency field}
\addcontentsline{toc}{subsection}{A stochastic replicator equation for the trait frequency field}

The dynamics of the per-capita population growth rate can be studied using Eq. \ref{density_SPDE}. By integrating Eq. \ref{density_SPDE} in the $x$ variable and dividing throughout by $N_K$, I find

\begin{linenomath*}\begin{equation}
		\frac{1}{N_K}\frac{dN_K}{dt} = \left[\overline{w}(t) + \mu \int\limits_{\mathcal{T}}Q(x|\phi)dx\right] + \frac{1}{\sqrt{KN_K(t)}}\int\limits_{\mathcal{T}}\sqrt{\phi(x,t) \tau(x|\phi) + \mu Q(x|\phi)}\dot{W}(x,t)dx
\end{equation}\end{linenomath*}
Thus, mean fitness controls the expected per-capita population growth rate, mean turnover rate controls the variance in the per-capita population growth rate, and mutations contribute to both mean and variance. Note that the stochastic term here is a simple purely temporal white noise rather than a spacetime white noise due to the integration over the $x$ variable. 

To describe evolutionary dynamics, we need an equation for trait frequencies. In supplementary section \ref{App_dens_to_freq}, I derive an SPDE for the trait frequency field using a heuristic infinite-dimensional It\^o formula~(\cite{curtain_itos_1970}, theorem 3.8; \cite{da_prato_stochastic_2014}, theorem 4.32). Given a type-level quantity $f[x|\phi]$, I define the \emph{selection-mutation} operator $\mathcal{S}_f[x|\phi]$ for $f$ in the population $\phi$ as:
\begin{linenomath*}\begin{equation}
		\mathcal{S}_f[x| \phi] = \underbrace{\vphantom{{\frac{\mu}{N_K(t)}\left( Q(x|\phi)-p(x)\int\limits_{\mathcal{T}} Q(y|\phi)dy\right)}}(f[x|\phi] - \overline{f}(t))p(x,t)}_{\substack{\text{Selection for higher}\\\text{values of $f$}}} + \underbrace{\frac{\mu}{N_K(t)}\left( Q(x|\phi)-p(x)\int\limits_{\mathcal{T}} Q(y|\phi)dy\right)}_{\text{Mutation biases/transmission biases}}
\end{equation}\end{linenomath*}
The operator $\mathcal{S}_f[x| \phi]$ represents how the trait frequency field $p$ changes at the point $x$ in a population $\phi$ as a balance between two evolutionary processes: (i) \emph{selection} for those trait values $x$ that are associated with higher values of $f$ than the population mean $\overline{f}$, and (ii) \emph{mutation} that can potentially bias which trait values $x$ arise in the population and thus how the trait frequency field changes over time. I show in section \ref{App_dens_to_freq} of the supplementary that the stochastic dynamics of the trait frequency field $p(x,t)$ are described by the remarkably compact equation:
\begin{linenomath*}\begin{equation}
		\label{freq_field}
		\frac{\partial p}{\partial t}(x,t) = \underbrace{\vphantom{\frac{1}{\sqrt{K}N_K(t)}}\mathcal{S}_w[x|\phi]}_{\substack{\text{Classical}\\\text{selection-}\\\text{mutation}}} - \underbrace{\vphantom{\frac{1}{\sqrt{K}N_K(t)}}\frac{1}{KN_K(t)}\mathcal{S}_{\tau}[x|\phi]}_{\substack{\text{Noise-induced}\\\text{selection-}\\\text{mutation}}} + \underbrace{\frac{1}{\sqrt{K}N_K(t)}\dot{W}_{p}(x,t)}_{\substack{\text{Undirected stochastic}\\\text{fluctuations}}}
\end{equation}\end{linenomath*}
where
\begin{linenomath*}\begin{equation}
		\label{freq_field_noise_term}
		\resizebox{\textwidth}{!}{$\displaystyle\dot{W}_{p}(x,t) \coloneqq \sqrt{\phi(x,t)\tau(x|\phi)+\mu Q(x|\phi)}\dot{W}(x,t)-p(x)\int\limits_{\mathcal{T}}\sqrt{\phi(y,t)\tau(y|\phi)+\mu Q(y|\phi)}\dot{W}(y,t)dy$}
\end{equation}\end{linenomath*}
is a spacetime white noise that vanishes upon taking probabilistic expectations.

Thus, the trait frequency field is influenced by three distinct evolutionary forces. $\mathcal{S}_w[x|\phi]$ quantifies the selection-mutation balance for higher fitness $w$ in the population and thus represents the effects of classical selection and mutation. The $-(\mathcal{S}_\tau[x|\phi]/KN_K)$ term quantifies the balance between mutation and selection for \emph{lower} turnover rates $\tau$ (notice the minus sign), and the strength of this force depends inversely on the total population size $KN_K$. This force is called noise-induced selection, and has been shown to play an important role in diverse finite population eco-evolutionary systems~\citep{constable_demographic_2016,mcleod_social_2019,week_white_2021,kuosmanen_turnover_2022,mazzolini_universality_2023,bhat_eco-evolutionary_2024}. Notice that even if a trait is in complete selection-mutation balance for Malthusian fitness $w$  (\emph{i.e.} $\mathcal{S}_w[x|\phi] = 0$), the trait frequency field could experience directional changes arising from a lack of selection-mutation balance for turnover rates $\mathcal{S}_{\tau}[x|\phi]$ due to noise-induced selection. Finally, $\dot{W}_{p}(x,t)$ captures the effects of stochastic fluctuations to the trait frequency field due to demographic stochasticity. This term exhibits the $1/\sqrt{K}$ scaling that is characteristic of genetic drift. Though the term disappears upon taking probabilistic expectations $\mathbb{E}[\cdot]$ and thus does not influence the expected behavior over short time scales, discrete trait analogs of this term are known to be able to directionally bias evolutionary dynamics over long timescales~\citep{mcleod_social_2019,mcleod_why_2019,bhat_eco-evolutionary_2024}. Equation \ref{freq_field} is a stochastic version of the replicator-mutator equation for quantitative traits in finite, fluctuating populations, as will become clear upon taking the infinite population limit.

\subsubsection*{The infinite population limit}

If we take the infinite population limit ($K \to \infty$) of equation \ref{freq_field}, all terms other than the selection-mutation operator for fitness drop out of the equation. Thus, the infinite population limit is the deterministic process described by the PDE
\begin{linenomath*}\begin{equation}
		\label{cts_replicator_mutator}
		\frac{\partial p}{\partial t}(x,t) = \left[w(x|\phi) - \overline{w}(t)\right]p(x,t)+\mu\left[Q(x|\phi) - p(x,t)\int\limits_{\mathcal{T}} Q(y|\phi)dy\right]
\end{equation}\end{linenomath*}
Equation \ref{cts_replicator_mutator} is a version of the replicator-mutator equation from evolutionary game theory for continuous strategy spaces~\citep{cressman_replicator_2014}. In supplementary section \ref{App_kimura}, I show that equation \ref{cts_replicator_mutator} also recovers Kimura's continuum-of-alleles model~\citep{kimura_stochastic_1965} when the trait space is the real line and the mutational effects in $Q(x|\phi)$ are modelled via convolution with a mutation kernel (\emph{i.e.} modelled such that mutations are symmetric, more extreme mutational effects are less likely, and the probability of a mutation of a given mutational effect size is parameterized by a mutation kernel function). The replicator-mutator equation can also be derived from a stochastic individual-based model using measure-theoretic martingale techniques~\citep{champagnat_unifying_2006,wakano_derivation_2017}.

\subsection*{A stochastic Price equation}
\addcontentsline{toc}{subsection}{A stochastic Price equation}

In supplementary section \ref{App_stoch_Price}, I use the equation for the trait frequency field (equation \ref{freq_field}) to show that the statistical mean value of any function $f$ (now possibly also varying over time) in the population obeys the one-dimensional SDE:

\begin{linenomath*}\begin{equation}
		\label{stoch_Price}
		\displaystyle\frac{d \overline{f}}{dt} = {\underbrace{\vphantom{\frac{dW_{\overline{f}}}{dt} \frac{1}{KN_K(t)} }\textrm{Cov}(w,f)}_{\substack{\text{Classical} \\ \text{selection}}}} - {\underbrace{\frac{1}{KN_K(t)}\textrm{Cov}(\tau,f)}_{\substack{\text{Noise-induced} \\ \text{selection}}}} +  {\underbrace{\vphantom{\frac{dW_{\overline{f}}}{dt} \frac{1}{KN_K(t)} }\overline{\left(\frac{\partial f}{\partial t}\right)}}_{\substack{\text{Ecological} \\ \text{effects}}}}
		+ \underbrace{\vphantom{\frac{dW_{\overline{f}}}{dt} \frac{1}{KN_K(t)}} M_{\overline{f}}(p,N_K)}_{\substack{\text{Mutational} \\ \text{effects}}}
		+ \underbrace{{\frac{1}{\sqrt{K}N_K(t)}\frac{dW_{\overline{f}}}{dt}}}_{\substack{\text{Stochastic} \\ \text{fluctuations}}}
\end{equation}\end{linenomath*}
where 
\begin{linenomath*}\begin{equation}
		M_{\overline{f}}(p,N_K) = \frac{\mu}{N_K}\left(1-\frac{1}{KN_K(t)}\right)\left(\int\limits_{\mathcal{T}} f(x|\phi) Q(x|\phi)dx-\overline{f}(t)\int\limits_{\mathcal{T}} Q(y|\phi)dy\right)
\end{equation}\end{linenomath*}
is a term describing the effects of mutation/transmission biases and
\begin{linenomath*}\begin{equation}
		\frac{dW_{\overline{f}}}{dt} = \int\limits_{\mathcal{T}} \left(f(x|\phi)-\overline{f}(t)\right)\sqrt{\phi(x,t)\tau(x|\phi)+\mu Q(y|\phi)}\dot{W}(x,t)dx
\end{equation}\end{linenomath*}
is a purely temporal white noise.

Eq. \ref{stoch_Price} once again reveals the effects of mutation and selection in a clear manner: The mean value of $f$ increases due to classical natural selection if $f$ covaries positively with fitness, and increases due to noise-induced selection if $f$ covaries negatively with turnover rate. Mutational biases are captured in the $M_{\overline{f}}$ term. The third term on the RHS of Eq. \ref{stoch_Price} is non-zero whenever the function $f$ changes over time through mechanisms other than through changes in the field $\phi$ itself, and thus represents the effects of eco-evolutionary feedbacks due to factors such as plasticity and environmental heterogeneity leading to changes in the function $f$ over time. This term also generically occurs in the Price equation and Fisher's fundamental theorem for discrete traits~\citep{lion_theoretical_2018,kokko_stagnation_2021}. Finally, the last term on the RHS of Eq. \ref{stoch_Price} represents the effects of undirected stochastic fluctuations that incorporate the effects of genetic/ecological drift. Note that Eq. \ref{stoch_Price} holds for any type level field $f(x|\phi)$. In supplementary section \ref{App_Fisher}, I present a stochastic analog of Fisher's fundamental theorem that arises upon substituting $f(x|\phi) = w(x|\phi)$ and $\mu = 0$ into Eq. \ref{stoch_Price}.

\subsubsection*{The infinite population limit}

If we take the infinite population limit in equation \ref{stoch_Price}, we obtain a deterministic ODE that reads
\begin{linenomath*}\begin{equation}
		\label{cts_price_general}
		\frac{d \overline{f}}{dt} = \mathrm{Cov}(w,f) + \overline{\left(\frac{\partial f}{\partial t}\right)} + \mu\left(\int\limits_{\mathcal{T}} f(x|\phi) Q(x|\phi)dx-\overline{f}(t)\int\limits_{\mathcal{T}} Q(y|\phi)dy\right)
\end{equation}\end{linenomath*}
Equation \ref{cts_price_general} is a (dynamic) version of the Price equation for quantitative traits. For the special case $f(x|\phi)=x$, the quantity $\partial f/\partial t$ is identically 0, and equation \ref{cts_price_general} reduces to a more familiar version of the Price equation~\citep{page_unifying_2002,lion_theoretical_2018}:
\begin{linenomath*}\begin{equation}
		\label{cts_price}
		\frac{d \overline{x}}{dt} = \mathrm{Cov}(w,x) + \mu\left(\int\limits_{\mathcal{T}} x Q(x|\phi)dx-\overline{x}(t)\int Q(y|\phi)dy\right)
\end{equation}\end{linenomath*}

\subsection*{A stochastic equation of gradient dynamics in finite populations}
\addcontentsline{toc}{subsection}{A stochastic equation of gradient dynamics in finite populations}

Consider now the special case $f(x|\phi) = x$ in Eq. \ref{stoch_Price}. In this section, I restrict myself to the strong selection, weak mutation limit. Specifically, I assume:
\begin{itemize}
	\item Rare mutations, \emph{i.e.} $\mu$ is infinitesimally small.
	\item Small mutational effects with `almost faithful' reproduction, meaning $Q(x|\phi)$ is infinitesimally small.
	\item Strong selection, meaning that types with low relative fitness are immediately eliminated and the population is sharply peaked around a few trait values. 
\end{itemize}
Mathematically, these assumptions mean that if we begin with a monomorphic population $\phi(x,0) = N_{K}(0)\delta_{y_0}$, the population (at least initially) remains strongly peaked about the mean value of the trait, with some small spread due to the (infinitesimal) mutational effects; Thus, I assume mathematically that $\sigma^2_x(t)$, the variance of the trait in the population at time $t$, is infinitesimal but non-zero. The density field $\phi(x,t)$ can then be approximated by a scaled Dirac delta mass $N_{K}(t)\delta_{y(t)}$ moving across the trait space according to a trajectory governed by a function $y(t)$ (to be found). I show in supplementary section \ref{App_gradient_dynamics} that under these assumptions, the trajectory $y(t)$ of a monomorphic population $\phi(x,t) = N_K(t)\delta_{y(t)}$ is approximately given by

\begin{linenomath*}
	\begin{equation}
		\label{stoch_gradient}
		\frac{dy}{dt} = \sigma^2_x(t) \underbrace{\frac{\partial G(x;y)}{\partial x}\bigg{|}_{x=y}}_{\substack{\text{Finite population}\\\text{Selection }\text{Gradient}}} + \underbrace{\vphantom{\frac{dG}{dx}(x;y)\bigg{|}_{x=y}}\frac{dW_{y}}{dt}}_{\substack{\text{Stochastic}\\\text{Fluctuations}}}
	\end{equation}
\end{linenomath*}
The quantity $G(x;y)$ is given by
\begin{linenomath*}
	\begin{equation}
		G(x;y) = \underbrace{\vphantom{\frac{1}{KN_K(t)}}w(x|N_K\delta_y)}_{\substack{\text{Classical}\\\text{Selection}}}-\underbrace{\frac{1}{KN_K(t)}\tau(x|N_K\delta_y)}_{\substack{\text{Noise-induced}\\\text{Selection}}}
	\end{equation}
\end{linenomath*}
and represents the balance between classical natural selection and noise-induced selection. $\partial G/\partial x$ is thus a modified selection gradient that not only accounts for classical selection, but also incorporates noise-induced selection. The white noise term in Eq. \ref{stoch_gradient} is given by
\begin{linenomath*}
	\begin{equation}
		\frac{dW_{y}}{dt}	= \int\limits_{\mathcal{T}}(x-y(t))\sqrt{\tau(x|N_K(t)\delta_{y(t)})}\dot{W}(x,t)dx
	\end{equation}
\end{linenomath*}
and vanishes upon taking probabilistic expectations over realizations. Note that the expected dynamics in the finite population do not follow the classic natural selection gradient as occurs in infinite population models, but instead follow a gradient that represents the balance between classical selection and noise-induced selection.~\citet{champagnat_evolution_2007} have also obtained an SDE for adaptive dynamics in finite populations that they call the `canonical diffusion' of adaptive dynamics. 

\subsubsection*{The infinite population limit}

Taking $K \to \infty$ in Eq. \ref{stoch_gradient}, we obtain
\begin{linenomath*}
	\begin{equation}
		\label{gradient_eqn}
		\frac{dy}{dt} = \sigma^2_x(t) \frac{\partial w(x|\delta_y)}{\partial x}\bigg{|}_{x=y}
	\end{equation}
\end{linenomath*}
The term $w\left(x|\delta_{y(t)}\right)$ is the expected growth rate of an individual with trait value $x$ in a population in which (almost) every individual has trait value $y$. This quantity is referred to as the invasion fitness of a `mutant' $x$ in a population of `resident' $y$ individuals. Eq. \ref{gradient_eqn} is the canonical form of a broad class of models captured under the name of `gradient dynamics'~\citep{abrams_relationship_1993, taylor_evolutionary_1997,lehtonen_price_2018}. It is also deeply related to the canonical equation of adaptive dynamics~\citep{lion_theoretical_2018,lehtonen_price_2018}.

\section*{Discussion}
\addcontentsline{toc}{section}{Discussion}

The stochastic field theoretic formalism I present provides a method for studying eco-evolutionary dynamics of populations bearing a single one-dimensional quantitative trait from the biological first principles of birth and death. In particular, I have derived an equation for studying ecological dynamics by tracking population densities (Eq. \ref{density_SPDE}) and have also derived SDE/SPDEs for evolutionary dynamics that generalize the replicator-mutator equation (Eq. \ref{freq_field}), Price equation (Eq. \ref{stoch_Price}), and gradient dynamics (Eq. \ref{stoch_gradient}). Along with describing the effects of natural selection, mutation/transmission bias, and genetic drift, these equations also provide a general description of the role of noise-induced selection in affecting the evolutionary trajectories of finite, fluctuating populations~\citep{gillespie_natural_1974,constable_demographic_2016,mcleod_social_2019,week_white_2021}. As a concrete example of the utility of these equations, I present a simple asexual model of resource competition in section~\ref{App_example} of the Supplementary that recovers the quantitative logistic equation~\citep{doebeli_adaptive_2011} in the infinite population limit, and a second example in~\ref{App_FKPP} that recovers as the Fisher-KPP equation in the infinite population limit but whose finite population dynamics does not correspond to the `stochastic Fisher-KPP equation'~\citep{doering_interacting_2003} (but is the SPDE expected by~\cite{champagnat_unifying_2006}).

One intriguing application of the general formalism outlined in this paper is in the study of the emergence of sympatric polymorphism for quantitative traits via evolutionary branching~\citep{doebeli_adaptive_2011}. Adaptive dynamics, the primary theoretical framework for studying evolutionary branching, is typically formulated in an infinite population setting~\citep{geritz_evolutionarily_1998,doebeli_adaptive_2011,avila_evolutionary_2023} obtained as a deterministic limit of an underlying stochastic model~\citep{dieckmann_dynamical_1996,champagnat_evolution_2007} . However, studies show that finite populations exhibit a systematically lower tendency to undergo evolutionary branching and/or take longer to branch than predicted by infinite population frameworks, and may remain monomorphic if the population size is too small~\citep{johansson_will_2006,claessen_delayed_2007,wakano_evolutionary_2013,debarre_evolutionary_2016,johnson_two-dimensional_2021}. SPDEs can often exhibit noise-induced phase transitions where stochasticity causes qualitative changes in the behavior of the system as a parameter controlling the strength of noise in the system is varied. Such transitions can be systematically studied using the language of non-equilibrium statistical physics~(\cite{garcia-ojalvo_noise_1999}, Chapter 3). The SPDEs I formulate in this paper, where the strength of stochastic fluctuations scales as $K^{-1/2}$, suggest that the failure of evolutionary branching in small populations could be reformulated and studied very generally in terms of a noise-induced phase transition in which the population size measure $K$ is the driving parameter. Alternatively, the stochastic version of the gradient equation (Eq. \ref{stoch_gradient}) could be used to study evolutionary branching in finite, fluctuating populations exactly analogously to how invasion fitness functions and their effects on population dynamics are used to study branching in infinite population models via the canonical equation of adaptive dynamics~\citep{doebeli_adaptive_2011}. In fact, Eq. \ref{stoch_gradient} directly shows that evolutionarily singular points (points at which the RHS of Eq. \ref{gradient_eqn} vanishes) need not be fixed points for finite populations when noise-induced selection is present. This fact could be a general factor hindering evolutionary branching in finite populations, since a population may not stay at an evolutionary branching point for long enough to allow polymorphisms to become established in the population. I also provide a general method to study evolutionary branching using the density field and a `weak noise approximation' in section \ref{sec_Fourier}.

\subsection*{Connections with previous studies}
\addcontentsline{toc}{subsection}{Connections with previous studies}

\citet{lande_natural_1976} has used tools from probability theory to study the effects of demographic stochasticity in populations bearing quantitative traits and evolving in discrete time. My work can be viewed as an extension of Lande's framework to fluctuating populations evolving in continuous time. Alternatively, the formalism can be seen as a generalization of~\citet{lion_theoretical_2018}'s conceptual synthesis of eco-evolutionary dynamics to finite, fluctuating populations: Taking the infinite population limit of the equations presented in this paper yields the quantitative traitversions of the equations presented in~\citet{lion_theoretical_2018}. Just like in~\citet{lion_theoretical_2018}, equations for moments such as the mean value (Eq. \ref{stoch_Price}) and variance of any field can be iteratively obtained from the stochastic replicator-mutator equation (Eq. \ref{freq_field}) using It\^o's formula.

The first study of noise-induced selection for reduced turnover rates is generally attributed to~\citet{gillespie_natural_1974}~\citep{veller_drift-induced_2017}. Gillespie was interested in variance in offspring numbers, and the effect I identify as noise-induced selection is therefore often referred to as a selection `for reduced variance' in the bet-hedging and life-history evolution literature since $\tau_i$ controls the infinitesimal variance of the density process through equation \ref{density_SPDE} (but note that the variance in this case is variance in per capita growth rate rather than variance in number of offspring, and that the stochasticity here is intrinsic to the population rather than being the result of a fluctuating external environment). My formulation can be connected with the bet-hedging literature in three distinct ways: (1) while \citet{gillespie_natural_1974} worked with simple models in which individuals could have only one of two possible phenotype values, the formalism I present in this paper provides equations describing noise-induced selection in populations bearing quantitative traits in which infinitely many distinct phenotypes may arise over time. (2) Later models of bet-hedging literature in life-history evolution often use an approach that is `dynamically insufficient': rather than finding dynamical equations that are forward-looking in time, these studies  instead partition a given amount of phenotypic change between two successive generations into various components in the style of the Price equation~\citep{frank_evolution_1990}. In contrast, my approach is dynamic, providing SDEs and SPDEs that are `forward-looking' in time. In this sense, the paper can be considered as a generalization of some of the ideas studied in~\cite{parsons_consequences_2010} and~\cite{bhat_eco-evolutionary_2024} to the study of quantitative traits. (3) The equations I provide in this paper also models a second effect of noise-induced selection that is often not discussed in bet-hedging literature: The `noise' terms (stochastic integral terms in the SDEs/SPDEs) have recently been shown to contribute to systematic directional biases in long-term evolutionary trajectories through a mechanism that is distinct from the `Gillespie effect' from bet-hedging theory ~\citep{mcleod_social_2019}. However, \citet{mcleod_social_2019} work with discrete traits, and my work can thus also be viewed as an extension of~\citet{mcleod_social_2019} to populations bearing quantitative traits.

\cite{week_white_2021} have recently independently arrived at the equations for trait frequencies (Eq. \ref{freq_field}) and mean trait value (Eq. \ref{stoch_Price} for the special case $f(x|\phi)=x$) by studying the scaling limits of measure-valued branching processes using certain heuristics for space-time white noise~\citep{week_white_2021}. My formalism and \cite{week_white_2021}'s formalism are complementary to each other. My formulation provides an alternate method of attack for the study of quantitative traits in finite fluctuating populations that may be more appropriate for some particular problems (such as phenotypic clustering; see section \ref{sec_Fourier}), while the approach in \cite{week_white_2021} may be more appropriate for others. ~\cite{etheridge_looking_2024} have also recently studied similar infinite-dimensional stochastic processes and their scaling limits in the context of spatial ecology. Though the focus of their work is ecological, many of the technical tools used and scaling limits studied are complementary to this paper as well as the more rigorous work of \cite{ champagnat_unifying_2006} and \cite{week_white_2021}.

\subsection*{The utility of the SPDE approach}
\addcontentsline{toc}{subsection}{The utility of the SPDE approach}

In this paper, I have used an analytical pipeline that consists of modelling the population as an infinite-dimensional birth-death process, describing the dynamics via a master equation, and then finding an approximate continuous approximation using a so-called `system size expansion' (Fig.~\ref{fig_summary}). This general approach first arose in statistical physics to describe the erratic motion of particles that are under the influence of a large number of forces~(\cite{kramers_brownian_1940,moyal_stochastic_1949,van_kampen_stochastic_1981}; for the infinite-dimensional version, see sections 13.1 and 13.2 of \cite{gardiner_stochastic_2009}). For discrete traits, the general analytical pipeline is well-known in population genetics, where it goes by the name of the `diffusion approximation'~\citep{feller_diffusion_1951,kimura_diffusion_1964}. Though most standard treatments of the diffusion approximation assume the total population size is strictly constant or varies deterministically~\citep{crow_introduction_1970,ewens_mathematical_2004, lambert_population_2010,czuppon_understanding_2021}, this assumption is not actually necessary~(\cite{feller_diffusion_1951}, section 10), and indeed, relaxing the assumption can have important consequences for population dynamics through noise-induced selection~\citep{gillespie_natural_1974,parsons_consequences_2010,constable_demographic_2016,mcleod_social_2019}.

I have shown how an approximation scheme that is very similar in spirit (Fig.~\ref{fig_summary}) can also be used to model the evolution of quantitative traits using functional derivatives and SPDEs rather than martingale techniques. My approach also helps clarify the mathematical connections between models of the evolutionary dynamics of discrete traits and those of quantitative traits --- informally, the field equations I present in this paper are the `$m \to \infty$' limit of equations describing the evolution of $m$ discrete traits upon replacing sums with integrals (see~\cite{bhat_eco-evolutionary_2024} for the discrete trait equations in the same notation as used in this paper). Conversely, discrete trait dynamics can be recovered from the field equations presented in this paper and in~\cite{week_white_2021} by discretizing the trait space, for instance by dividing the trait space $\mathcal{T} \subset \mathbb{R}$ into $m$ disjoint intervals and treating all individuals that have trait values within the same interval as equivalent (Fig.~\ref{fig_summary}).

The equations I derive in this paper also allow us to leverage tools from dynamical systems that complement the tools that come with the more probabilistic approach used in the current literature. For instance, models in theoretical population biology and population genetics routinely assume a separation between ecological and evolutionary timescales~\citep{parsons_consequences_2010,constable_stochastic_2013,chotibut_population_2017,mcleod_social_2019} to make stochastic dynamics more amenable to analysis.~\citet{parsons_dimension_2017} have recently extended the relevant mathematical machinery (`adiabatic elimination' in physics language, `slow manifold approximation' in mathematics language) to infinite dimensional systems (i.e. SPDEs). The SPDEs I present in this paper may thus allow us to by-pass the formidable stochastic analysis tools that are required for formulating timescale separation arguments in the more rigorous measure-theoretic martingale perspective~(see, for instance, section 5 in \cite{champagnat_individual_2008}). SPDEs are also more amenable for studying noise-induced oscillations in population abundance~(\cite{garcia-ojalvo_noise_1999}, Chapter 5) using spectral methods similar to those in supplementary section \ref{sec_Fourier}.

Currently, stochastic field equations of the kind I derive here are primarily used by statistical physicists~\citep{garcia-ojalvo_noise_1999} and are attacked using ingenious heuristic tools such as the path integral formalism~\citep{hochberg_effective_1999, chow_path_2015, weber_master_2017}, Feynman diagrams~\citep{thomas_system_2014}, Fock space methods~\citep{del_razo_probabilistic_2022}, and the renormalization group~\citep{tauber_applications_2005}. The equations and general approach I develop are also intended to encourage the use of such techniques from physics in studying the evolution of quantitative traits. 

\myfig{0.9}{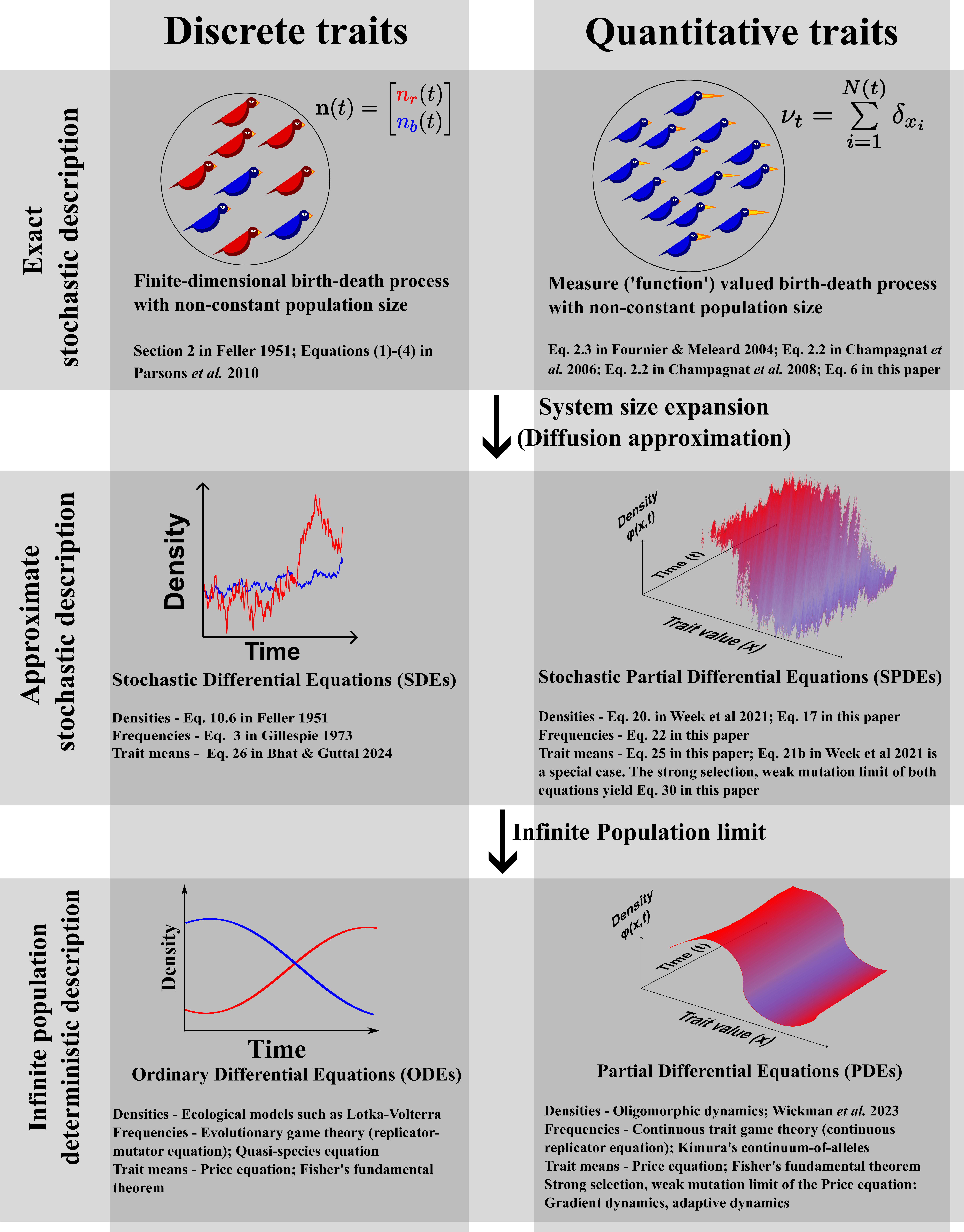}{\textbf{Summary of the analytic pipeline used in this manuscript and a comparison with discrete trait models.} Papers listed provide examples of previous studies that are situated at various points in this pipeline. I have deliberately tried to cite older papers wherever possible. \cite{page_unifying_2002}, \cite{lion_theoretical_2018}, and \cite{lehtonen_price_2018} speak about the bottom most panels and the connections between the various deterministic equations.~\cite{champagnat_unifying_2006} discuss the rightmost panel from the measure-theoretic perspective.}{fig_summary}

\subsection*{Summary and Outlook}
\addcontentsline{toc}{subsection}{Summary and Outlook}

In this paper, I have presented a field theoretic approach to modeling the eco-evolutionary population dynamics of quantitative traits in finite, fluctuating populations. The equations I derive provide a generic description of evolutionary dynamics in finite, fluctuating populations that includes the effects of noise-induced selection alongside the more standard forces of natural selection, mutation, eco-evolutionary feedbacks, and genetic/ecological drift. My approach uses techniques grounded in statistical physics and the calculus of variations that are analogous to the diffusion approximation from population genetics~(Fig.~\ref{fig_summary}) and presents an alternative perspective that complements current formulations of evolutionary dynamics of quantitative traits in finite populations that typically use martingale techniques~\citep{champagnat_unifying_2006,champagnat_individual_2008,boussange_eco-evolutionary_2022,etheridge_looking_2024}. 

Importantly, the formalism I develop here likely does \emph{not} carry over to the study of population dynamics in higher dimensional trait spaces. This is because stochastic processes driven by spacetime white noise are habitually badly behaved in higher spatial dimensions, making analytical progress very difficult. For example, measure-valued birth-death processes and their scaling limits (superprocesses) often do not admit a density with respect to the Lebesgue measure in $\geq 2$ dimensions~(\cite{fleming_measure-valued_1979,konno_stochastic_1988, dawson_stochastic_2000,etheridge_introduction_2000}; also see Remark 4.1.1.2.3 in~\cite{champagnat_individual_2008} and Remark 2.20 in~\cite{etheridge_looking_2024}). Even if we were to ignore this technical point, SPDEs in $\geq 2$ dimensions routinely do not even admit any function valued solutions~\citep{etheridge_introduction_2000,pardoux_stochastic_2021} and are thus difficult to handle analytically. It may well be the case that concrete biologically useful progress in this direction requires radically new mathematics, a situation increasingly also encountered in other areas of mathematical biology~\citep{borovik_mathematicians_2021,vittadello_open_2022}.

\section*{Acknowledgements}

This paper arose from work conducted during my Master's thesis project in Vishwesha Guttal's group at the Centre for Ecological Sciences (CES) at the Indian Institute of Science. During that time,  Prof. Guttal's group was supported by grants from the Science and Engineering Research Board (SERB), Government of India (MTR/2022/000273). I am grateful to Vishwesha Guttal for his guidance, advice, and several insightful discussions. Gaurav Athreya, Anish Bhattacharya, Peter Ralph, and two anonymous reviewers provided helpful comments that greatly improved an earlier draft of this manuscript. I am a recipient of a Kishore Yaigyanik Protsahan Yojana (KVPY) fellowship from the Department of Science and Technology of the Government of India (Fellowship ID: SX-1711025), and acknowledge the same for financial support. The open access fee for this publication was covered by the DEAL Konsortium between Elsevier and Johannes Gutenberg University, Mainz.

\phantomsection
\addcontentsline{toc}{section}{References}
\printbibliography[title=References]
\end{refsection}

\clearpage

\setcounter{page}{1} 

\newpage

\begin{center}
	{\LARGE \bfseries Supplementary Information for \\
		Bhat 2024: A stochastic field theory for the evolution of quantitative traits in finite populations\\
		\vspace{2em}
		\normalfont\Large Ananda Shikhara Bhat\textsuperscript{1,2,*}\\
		\vspace{3em}
		\large \textsuperscript{1}Dept. of Biology, Indian Institute of Science Education and Research, Pune, India\\
		\textsuperscript{2}Centre for Ecological Sciences, Indian Institute of Science, Bengaluru, India\\
		\vspace{1em}
		\textsuperscript{*} \textit{Current Affiliation:} Institute of Organismic and Molecular Evolution (iomE) and Institute for Quantitative and Computational Biosciences (IQCB), Johannes Gutenberg University, 55128 Mainz, Germany. E-mail for correspondence: \href{mailto:abhat@uni-mainz.de}{abhat@uni-mainz.de}}
\end{center}
\newpage
\clearpage
\setcounter{equation}{0}  
\counterwithout{equation}{section}

\renewcommand{\theequation}{S\arabic{equation}}
\renewcommand{\thetable}{S\arabic{table}}
\renewcommand{\thesection}{S\arabic{section}}
\renewcommand{\thefigure}{S\arabic{figure}}
\setcounter{figure}{0}
\setcounter{table}{0}
\setcounter{section}{0}

\pagestyle{fancy}
\rhead{Supplement to Bhat 2024, 
	``Evolutionary field theory''}
\setlength{\headsep}{0.4in}  
\lhead{} 
\setlength{\headheight}{15pt}

\linespread{1.2}

\begin{refsection}
\startcontents[TOC]

\newpage

\begin{center}
	{\Large\bfseries Contents\\}
	\vspace{2em}
\end{center}

\printcontents[TOC]{l}{1}[3]{}

\newpage

\section{The Big Picture}\label{App_outline}

In this document, I provide detailed derivations of the equations discussed in the main text. I use this section to provide a verbal summary of the motivation behind the calculations I carry out in subsequent sections. To study ecological dynamics, we would like an equation that describes how population densities change over time. While the functional master equation, Eq. \ref{maintext_unnormalized_M_equation}, in principle describes the entire population at an exact level, in practice it is usually much too complicated to be studied directly. Much of the complication is due to the discontinuity inherent in discrete population `jumps', and so an approximate continuous description is desirable. In section \ref{App_system_size}, I use an infinite-dimensional version of the system-size expansion~\citep{gardiner_stochastic_2009} (`diffusion approximation' in population genetics) to derive an approximate functional Fokker-Planck equation for the stochastic dynamics of the population density field.

The basic idea of a system-size approximation originates in physics~(\cite{van_kampen_stochastic_1981}, chapter 10), where it was observed that when particles (or molecules) move and interact stochastically in a container of volume $\Omega$, the transitions in the state of the system are in terms of changes to particle number $n \to n \pm 1$ through the creation/annihilation of particles, but the rate of these transitions is governed by the rate at which particles meet, which is proportional to the density $n/\Omega$ of particles per unit volume. When viewed in units of particle density $n/\Omega$, the system transitions in units of $1/\Omega$ and thus looks approximately continuous if $\Omega$ is not too small. In our biological language, birth rates, death rates, and interactions are in terms of population density, whereas the system itself is described in terms of the number of individuals. The two are related by an abstract `population size measure'~\citep{czuppon_understanding_2021} that plays the role of the system size parameter $\Omega$. In supplementary section~\ref{App_system_size}, I carry out an infinite-dimensional analog of this idea to derive a functional Fokker-Planck equation which I present as Eq.~\ref{density_FPE} in the main text. This equation is equivalent to a stochastic partial differential equation (SPDE), namely Eq.~\ref{density_SPDE} in the main text. 

To study evolutionary dynamics, we need to derive an equation for how trait frequencies change. When dynamics are stochastic, one cannot use the usual formulae for change of variables. Instead, we require a tool from stochastic calculus called It\^o's formula. In supplementary section \ref{App_dens_to_freq}, I use Eq.~\ref{density_FPE} along with an infinite-dimensional version of It\^o's formula~(\cite{curtain_itos_1970}, theorem 3.8; \cite{da_prato_stochastic_2014}, theorem 4.32) to derive an SPDE for the trait frequency field $p(x,t)$ that generalizes the replicator-mutator equation to finite, fluctuating populations bearing quantitative traits. This results in Eq.~\ref{freq_field} in the main text. In section~\ref{kimura_continuum_model}, I illustrate how Eq.~\ref{freq_field} recovers Kimura's continuum-of-alleles model in the infinite population limit to underscore the fact that my equation is an extension of known models in population biology.

Studying SPDEs can be complicated and the tools to analyze them are still in development. On the other hand, stochastic differential equations (SDEs) are much more well-understood. For practical applications, it may thus be better to derive an SDE to describe (at least some aspects of) the behavior of the frequency field $p(x,t)$ via some aggregate quantity. A natural candidate is the mean trait value $\overline{x}$ in the population, or more generally, the mean value $\overline{f}$ of any function $f(x|\phi)$ of the trait value $x$ in the population $\phi$. In section~\ref{App_stoch_Price}, I use the equation for the trait frequency field derived in section~\ref{App_dens_to_freq} to derive an SDE for the mean value of any function $f$ of the trait value. This equation ends up being a stochastic generalization of the celebrated Price equation, and is Eq.~\ref{stoch_Price} in the main text. Fisher's fundamental theorem, a corollary of the Price equation~\citep{queller_fundamental_2017}, is an equation for the mean fitness $\overline{w}$ in the population and has historical significance. As an illustration of the stochastic Price equation Eq.~\ref{stoch_Price}, I derive a stochastic version of Fisher's fundamental theorem in section~\ref{App_Fisher}. I also illustrate that the resultant SDE indeed recovers the standard version of Fisher's fundamental theorem in the infinite population limit.

Under certain scaling limits, the Price equation (in continuous time and the standard infinite population setting) heuristically recovers Lande's gradient equation~\citep{page_unifying_2002,lehtonen_price_2018}, an intuitively appealing equation describing evolution as following a gradient in fitness. Recovering the gradient equation is also of interest because it is closely related to the canonical equation of adaptive dynamics~\citep{page_unifying_2002,lion_theoretical_2018,lehtonen_price_2018}, the fundamental machinery behind the enormous field of adaptive dynamics~\citep{doebeli_adaptive_2011}. In section~\ref{App_gradient_dynamics}, I illustrate how similar heuristic arguments can also be used in our stochastic setting to recover a stochastic version of the gradient equation (Eq.~\ref{stoch_gradient} in the main text). Strikingly, the stochastic gradient equation we derive does \emph{not} predict that evolution follows a gradient for higher fitness alone, but instead also predicts a detectable effect of \emph{noise-induced selection}, a phenomenon particular to finite populations. Nevertheless, our equation recovers the standard gradient equation in the infinite population limit.

What good is all this abstract theory? Do we gain anything that we would not already have known with the rigorous measure-theoretic martingale framework of papers such as~\citet{champagnat_unifying_2006}? Besides being accessible to a different audience, I believe the SPDE approach is also more suited to studying certain biological questions (while the more probabilistic approach of~\citet{champagnat_unifying_2006} is better suited for others) and thus presents a useful complementary perspective on the evolution of quantitative traits in finite populations. I illustrate this point in section~\ref{sec_Fourier} by using spectral methods and linear approximations to study finite-population effects on phenotypic clustering and adaptive diversification over long timescales. This section presents a general treatment of techniques that have been used to study particular ecological problems in previous studies~\citep{rogers_demographic_2012,rogers_modes_2015}. Another concrete example of similar complementary approaches to studying stochastic processes is seen in Wright-Fisher dynamics, where ideas from stochastic calculus~\citep{chen_fundamental_2010} and partial differential equations~\citep{epstein_wrightfisher_2010} have been used in complementary fashion in parallel to attack the same evolutionary question through different routes.

Finally, sections~\ref{App_example} and~\ref{App_FKPP} provide simple examples to illustrate how the entire analytic pipeline can be put to use. More specifically, I start with a simple Lotka-Volterra style resource competition model at the individual (`microscopic') level and show how the `mesoscopic' behavior yields SPDEs for the popualtion densities and trait frequencies. In the infinite population (`macroscopic') limit, these equations recover the quantitative logistic equation in the population density case, and a standard model from adaptive dynamics in the gradient equation case. In the Fisher-KPP case, I show that starting from an individual-based birth-death process, the `mesoscopic' limit is \emph{not} the SPDE usually referred to as the `stochastic Fisher-KPP equation'~\citep{doering_interacting_2003,barton_modelling_2013} but instead the SPDE expected by~\citet{champagnat_unifying_2006}. The purpose of this section is once again to illustrate how well-known models can be recovered from `microscopic' first principles in our framework.

\section{A functional system-size expansion to obtain the equation for the population density field}\label{App_system_size}
I begin with the master equation, Eq. \ref{maintext_unnormalized_M_equation}:
\begin{linenomath*}\begin{equation}
		\label{unnormalized_M_equation}
		\frac{\partial P}{\partial t}(\nu,t) = \int\limits_{\mathcal{T}}\left[(\mathcal{E}^{-}_{x}-1)b(x|\nu)P(\nu,t) + (\mathcal{E}^{+}_{x}-1)d(x|\nu)P(\nu,t)\right]dx
\end{equation}\end{linenomath*}
As mentioned in the main text, I introduce the population density field $\phi$ via the transformation
\begin{linenomath*}\begin{equation*}
		\phi(\cdot,t) \coloneqq \frac{1}{K}\nu(\cdot,t) = \frac{1}{K}\sum\limits_{i=1}^{N(t)}\delta_{x_i}
\end{equation*}\end{linenomath*}
and assume we can find $\mathcal{O}(1)$ functions $b_K$ and $d_K$ such that the original birth and death rate functions $b$ and $d$ can be rewritten as:
\begin{linenomath*}\begin{equation}
		\label{pop_density_BD_variable_transform}
		\begin{aligned}
			b(x|\nu) &= Kb_K(x|\nu/K) = Kb_K(x|\phi)\\
			d(x|\nu) &= Kd(x|\nu/K) = Kd_K(x|\phi)
		\end{aligned}
\end{equation}\end{linenomath*}
where $b_K$ and $d_K$ are $\mathcal{O}(1)$. In terms of these new variables, Eq. \ref{unnormalized_M_equation} becomes:
\begin{linenomath*}\begin{equation}
		\label{M_equation}
		\frac{\partial P}{\partial t}(\phi,t) = K\int\limits_{\mathcal{T}}\left[(\Delta^{-}_{x}-1)b_K(x|\phi)P(\phi,t) +(\Delta^{+}_{x}-1)d_K(x|\phi)P(\phi,t)\right]dx
\end{equation}\end{linenomath*}
where I have introduced new step operators $\Delta_{x}^{\pm}$ that satisfy:
\begin{linenomath*}\begin{equation*}
		\Delta_{x}^{\pm}[F(y,\phi)] =  F\left(y,\phi \pm \frac{1}{K}\delta_x\right)
\end{equation*}\end{linenomath*}

We can now conduct a system-size expansion  by using a functional analog of a Taylor expansion of the step operators. Below, I suppress the $t$ dependence of $\phi$ for notational conciseness. Recall that the functional version of the Taylor expansion of a functional $F[\rho]$ about a function $\rho_0$ defined on a domain $\Omega \subseteq \mathbb{R}$ is given by:

\begin{linenomath*}\begin{equation*}
		F[\rho_0 + \rho] = F[\rho_0] + \int\limits_{\Omega}\rho(x)\frac{\delta F}{\delta \rho_0(x)}dx + \frac{1}{2!}\int\limits_{\Omega}\int\limits_{\Omega}\rho(x)\rho(y)\frac{\delta^2 F}{\delta \rho_0(x)\delta \rho_0(y)}dxdy + \cdots
\end{equation*}\end{linenomath*}

Since $\Delta^{\pm}_{x}[F[\phi]] = F[\phi \pm \delta_x/K]$, we can Taylor expand the RHS to see that our step operators obey
\begin{linenomath*}\begin{align}
		\Delta^{\pm}_{x}[F[\phi]] &= F[\phi] \pm \frac{1}{K}\int\limits_{\mathcal{T}}\frac{\delta F}{\delta \phi(y)}\delta_xdy + \frac{1}{2K^2}\int\limits_{\mathcal{T}}\int\limits_{\mathcal{T}}\frac{\delta^2 F}{\delta \phi(y)\delta \phi(z)}\delta_xdy\delta_xdz+\cdots\\
		&= F[\phi] \pm \frac{1}{K}\frac{\delta F}{\delta \phi(x)} + \frac{1}{2K^2}\frac{\delta^2 F}{\delta \phi(x)^2}+\cdots
		\label{KM_ansatz}
\end{align}\end{linenomath*}
We will now neglect all higher order terms. Mathematically, this is justified by Pawula's theorem~\citep{pawula_approximation_1967}. Biologically, neglecting all higher order terms amounts to saying that the stochastic field $\phi$ is entirely described by its first two moments and thus is a Gaussian approximation~\citep{black_stochastic_2012}. Neglecting the higher order terms, we can now substitute Eq. \ref{KM_ansatz} with $F = b_K(x|\phi)P(\phi,t)$ and $F = d_K(x|\phi)P(\phi,t)$ into Eq. \ref{M_equation} to obtain:
\begin{linenomath*}\begin{equation*}
		\begin{split}
			\frac{\partial P}{\partial t}(\phi,t) = K\int\limits_{\mathcal{T}}\left[
			\left(-\frac{1}{K}\frac{\delta}{\delta\phi(x)} + \frac{1}{2K^2}\frac{\delta^2}{\delta\phi(x)^2}\right)\{b_K(x|\phi)P(\phi,t)\}\right]dx\\
			+K\int\limits_{\mathcal{T}}\left[\left(\frac{1}{K}\frac{\delta}{\delta\phi(x)} + \frac{1}{2K^2}\frac{\delta^2}{\delta\phi^2(x)}\right)\{d_K(x|\phi)P(\phi,t)\}\right]dx
		\end{split}
\end{equation*}\end{linenomath*}
Rearranging these terms, we obtain a `functional Fokker-Planck equation':
\begin{linenomath*}\begin{equation}
		\label{functional_FPE}
		\setlength{\fboxsep}{2\fboxsep}\boxed{\frac{\partial P}{\partial t}(\phi,t) = \int\limits_{\mathcal{T}}\left[-
			\frac{\delta}{\delta\phi(x)}\{\mathcal{A}^{-}(x|\phi)P(\phi,t)\} + \frac{1}{2K}\frac{\delta^2}{\delta\phi(x)^2}\{\mathcal{A}^{+}(x|\phi)P(\phi,t)\}\right]dx}
\end{equation}\end{linenomath*}
where
\begin{linenomath*}\begin{align*}
		\mathcal{A}^{\pm}(x|\phi) &= b_K(x|\phi)\pm d_K(x|\phi) = \frac{1}{K}\left(b(x|\nu)\pm d(x|\nu)\right)
\end{align*}\end{linenomath*}
Equation \ref{functional_FPE} yields Eq. \ref{density_FPE} in the main text upon substituting the functional forms of birth and death rates given in Eq. \ref{BD_defns}.

\section{Deriving stochastic trait frequency dynamics via a heuristic It\^o formula}\label{App_dens_to_freq}

In this section, I derive an SPDE for the trait frequency field. I will do this by assuming an `intuitive' infinite-dimensional It\^o formula holds for SPDEs. The formula in question is described below.

Let $\mathcal{T} \subset \mathbb{R}$, and let $\mathcal{M}(\mathcal{T})$ be a suitable space of functions from $\mathcal{T}$ to $\mathbb{R}$. Let $\mathcal{F}[\phi,x]:\mathcal{M}(\mathcal{T}) \times \mathcal{T} \to \mathbb{R}$ and $\mathcal{G}[\phi,x,y]:\mathcal{M}(\mathcal{T}) \times \mathcal{T} \times \mathcal{T} \to \mathbb{R}$ be two functionals. Consider the spacetime stochastic process $\phi(x,t)$ obtained as the solution to the SPDE
\begin{linenomath*}\begin{equation}
		\label{general_SPDE}
		\frac{\partial \phi}{\partial t} (x,t) = \mathcal{F}[\phi,x] + \int\limits_{\mathcal{T}}\mathcal{G}[\phi,x,y]\dot{W}(y,t)dy
\end{equation}\end{linenomath*}
Let $\mathcal{H}:\mathcal{M}(\mathcal{T}) \to \mathbb{R}$ be any `nice' functional. Then, I assume that $\mathcal{H}[\phi]$ satisfies the integral equation
\begin{linenomath*}\begin{equation}
		\label{general_Ito_integral_eqn}
		\begin{split}
			\mathcal{H}[\phi(\cdot,t)] &= \mathcal{H}[\phi(\cdot,0)] + \int\limits_{0}^{t} \int\limits_{\mathcal{T}}\mathcal{F}[\phi(\cdot,s),x]\frac{\delta \mathcal{H}}{\delta \phi(x,s)}dxds\\
			&+ \frac{1}{2}\left(\int\limits_{0}^{t}\int\limits_{\mathcal{T}}\int\limits_{\mathcal{T}} \frac{\delta^2\mathcal{H}}{\delta\phi(x,s)\delta\phi(y,s)}\left(\int\limits_{\mathcal{T}}\mathcal{G}[\phi(\cdot,s),x,z]\mathcal{G}[\phi(\cdot,s),y,z]dz\right)dxdyds\right)\\
			&+ \int\limits_{0}^{t}\int\limits_{\mathcal{T}}\int\limits_{\mathcal{T}} \mathcal{G}[\phi(\cdot,s),x,y]\frac{\delta \mathcal{H}}{\delta \phi(x,s)}\dot{W}(y,s)dxdyds
		\end{split}
\end{equation}\end{linenomath*}
Here, I have used the notation $\phi(\cdot,s)$ to refer to the entire density field $\phi$ at time $s$, to be distinguished from  $\phi(x,s)$, the particular value of the density field at the trait value $x$ at time $s$. In other words, $\phi(\cdot,s)$ is a function (element of $\mathcal{M}(\mathcal{T})$) whereas $\phi(x,s)$ is a real number.

In SDE/SPDE notation, equation Eq. \ref{general_Ito_integral_eqn} can be written in the form of an (infinite-dimensional) It\^o formula as:
\begin{linenomath*}\begin{equation}
		\label{heuristic_Ito}
		\resizebox{\textwidth}{!}{
			$\displaystyle
			\begin{split}
				d\mathcal{H}[\phi(\cdot,t)] &= \left[ \int\limits_{\mathcal{T}}\mathcal{F}[\phi(\cdot,t),x]\frac{\delta \mathcal{H}}{\delta \phi(x,t)}dx + \frac{1}{2}\int\limits_{\mathcal{T}}\int\limits_{\mathcal{T}} \frac{\delta^2\mathcal{H}}{\delta\phi(x,t)\delta\phi(y,t)}\left(\int\limits_{\mathcal{T}}\mathcal{G}[\phi(\cdot,t),x,z]\mathcal{G}[\phi(\cdot,t),y,z]dz\right)dxdy\right]dt\\
				& + \int\limits_{\mathcal{T}}\int\limits_{\mathcal{T}} \mathcal{G}[\phi(\cdot,t),x,y]\frac{\delta \mathcal{H}}{\delta \phi(x,t)}\dot{W}(y,t)dxdy
			\end{split}
			$
		}
\end{equation}\end{linenomath*}
Note that equation Eq. \ref{heuristic_Ito} is precisely the expression we would intuitively expect to obtain if we informally ``take $n\to\infty$'' in the $n$-dimensional It\^o formula for finite $n$~\citep{oksendal_stochastic_1998,gardiner_stochastic_2009}. Such an It\^o formula is rigorously known to hold in various special cases when the domain $\mathcal{T}$, the function space $\mathcal{M}(\mathcal{T})$, and the functionals (operators) $\mathcal{F},\mathcal{G},\mathcal{H}$ all satisfy certain technical assumptions~(\cite{curtain_itos_1970}, theorem 3.8; \cite{dawson_stochastic_1975}, theorem 4.12; \cite{da_prato_stochastic_2014}, theorem 4.32; \cite{week_white_2021}, section SM2). I will make no attempt to characterize whether our system satisfies these technical assumptions and will simply assume that equation Eq. \ref{heuristic_Ito} holds for our purposes.

To derive an expression for the trait frequency field at time $t$, let us define for each $x \in \mathcal{T}$ a functional
\begin{linenomath*}\begin{equation*}
		\mathcal{H}_x[\rho] = \frac{\rho(x)}{\int\limits_{\mathcal{T}}\rho(u)du}
\end{equation*}\end{linenomath*}
Note that when applied on the population density field $\phi$, $\mathcal{H}_x$ returns the value of the trait frequency field at the point $x$.
Now, let $\phi$ denote the population density field, given by the solution to the SPDE described by Eq.~\ref{functional_FPE} (the SPDE is Eq.~\ref{density_SPDE} in the main text). Comparing terms with Eq. \ref{general_SPDE}, we can identify
\begin{linenomath*}\begin{align}
		\mathcal{F}[\phi,x] &= \phi(x,t) w(x|\phi) + \mu Q(x|\phi)\\
		\mathcal{G}[\phi,x,y] &= \frac{1}{\sqrt{K}}\left[\left(\phi(x,t) \tau(x|\phi) + \mu Q(x|\phi)\right)\left(\phi(y,t) \tau(y|\phi) + \mu Q(y|\phi)\right)\right]^{1/4}\delta_{x,y}
\end{align}\end{linenomath*}
where $\delta_{x,y}$ is the (generalized) Kronecker delta, defined as
\begin{linenomath*}\begin{equation*}
		\delta_{x,y} \coloneqq \begin{cases} 1 & x=y\\
			0 & x\neq y
		\end{cases}
\end{equation*}\end{linenomath*}
Thus, by our It\^o formula Eq. \ref{heuristic_Ito}, $\mathcal{H}_x[\phi]$ (and hence the trait frequency field at point $x$) satisfies
\begin{linenomath*}\begin{equation}
		\label{density_Ito}
		\begin{split}
			d\mathcal{H}_{x}[\phi] &= \bigg{[} \int\limits_{\mathcal{T}}\left\{\phi(y,t)w(y|\phi)+\mu Q(y|\phi)\right\}\frac{\delta \mathcal{H}_x}{\delta \phi(y,t)}dy\\
			&\hphantom{=}+ \frac{1}{2K}\left(\int\limits_{\mathcal{T}}  \left[\phi(y,t)\tau(y|\phi)+\mu Q(y|\phi)\right]\frac{\delta^2 \mathcal{H}_x}{\delta\phi(y,t)^2}dy\right)\bigg{]}dt\\
			&\hphantom{=}+\frac{1}{\sqrt{K}}\int\limits_{\mathcal{T}} \sqrt{\phi(y,t)\tau(y|\phi)+\mu Q(y|\phi)}\frac{\delta \mathcal{H}_x}{\delta \phi(y,t)}\dot{W}(y,t)dy
		\end{split}
\end{equation}\end{linenomath*}

We will evaluate the RHS of Eq. \ref{density_Ito} term by term.

To begin with, let us calculate the two functional derivatives that appear on the RHS of Eq. \ref{density_Ito}. To do this, we will use the identity

\begin{linenomath*}\begin{equation}
		\label{generic_functional_derivative_identity}
		\frac{\delta F}{\delta \rho(x)} = \lim\limits_{\epsilon \to 0}\frac{F[\rho + \epsilon\delta_x]-F[\rho]}{\epsilon}
\end{equation}\end{linenomath*}
obtained by substituting $\xi(x) = \delta_x$ in the definition Eq. \ref{functional_derivative_defn} of the functional derivative. Note that the use of a delta mass as a test function is justified for our purposes because, in our cases, $F$ will be defined on $\mathcal{M}_K(\mathcal{T})$, and thus, from the definition of $M_K(\mathcal{T})$, if $x \in \mathcal{T}$ and $\epsilon > 0$, $\rho + \epsilon \delta_x$ will be in the domain of $F$ provided $\rho$ is in the domain of $F$. For the rest of this section, I use the notational shorthand $\int = \int_{\mathcal{T}}$ and suppress the $t$ dependence of all fields for conciseness.

Let $y \in \mathcal{T}$. We can calculate the single functional derivative as
\begin{linenomath*}\begin{align}
		\frac{\delta\mathcal{H}_x}{\delta\phi(y)} &= \lim\limits_{\epsilon \to 0}\frac{1}{\epsilon}\left[\mathcal{H}_x[\phi + \epsilon\delta_{\cdot}]-\mathcal{H}_x[\phi]\right]\\
		&= \lim\limits_{\epsilon \to 0}\frac{1}{\epsilon}\left[\frac{\phi(x)+\epsilon\delta_x}{\int\phi(u)du+\epsilon\int\delta_udu}-\frac{\phi(x)}{\int\phi(u)du}\right]\\
		&=  \lim\limits_{\epsilon \to 0}\frac{1}{\epsilon}\left[\frac{\epsilon\left(\delta_x\int\phi(u)du-\phi(x)\right)}{[\int\phi(u)du]^2+\epsilon\int\phi(u)du}\right]\\
		&= \frac{1}{\int\phi(u)du}\left[\delta_x-p(x)\right]\label{for_freq_first_derivative}
\end{align}\end{linenomath*}
where I have used the notation $p(x) = \phi(x)/\int\phi(u)du$ for the trait frequency field from the main text Eq. \ref{freq_defn}.
Now, let $z \in \mathcal{T}$. We can rewrite the double functional derivative in equation Eq. \ref{density_Ito} as
\begin{linenomath*}\begin{equation}
		\frac{\delta^2 \mathcal{H}_x}{\delta \phi(y)^2} = \frac{\delta}{\delta \phi(y)}\left(\frac{\delta\mathcal{H}_x}{\delta\phi(y)}\right) = \frac{\delta \mathcal{H}'_x[\phi]}{\delta \phi(y)}
\end{equation}\end{linenomath*}
where, from Eq. \ref{for_freq_first_derivative}, we know
\begin{linenomath*}\begin{equation}
		\mathcal{H}'_x[\phi] = \frac{1}{\int\phi(u)du}\left(\delta_x-\frac{\phi(x)}{\int\phi(u)du}\right)
\end{equation}\end{linenomath*}
Using Eq. \ref{generic_functional_derivative_identity}, we now obtain
\begin{linenomath*}
	\begin{equation}
		\resizebox{\textwidth}{!}{
			$\displaystyle
			\frac{\delta^2 \mathcal{H}_x}{\delta \phi(y)^2} =  \lim\limits_{\epsilon \to 0}\frac{1}{\epsilon}\left[\frac{\delta_x}{\int\phi(u)du+\epsilon\int\delta_udu}-\frac{\phi(x)+\epsilon\delta_x}{(\int\phi(u)du+\epsilon\int\delta_udu)^2}-\left(\frac{\delta_x}{\int\phi(u)du}-\frac{\phi(x)}{(\int\phi(u)du)^2}\right)\right]
			$
		}
	\end{equation}
\end{linenomath*}
\begin{linenomath*}\begin{align}
		&= \lim\limits_{\epsilon \to 0}\frac{1}{\epsilon}\left[\frac{-\epsilon\delta_x}{\int\phi(u)du+\epsilon}+\frac{(\int\phi(u)du)^2(\phi(x)+\epsilon\delta_x)+\phi(x)(\int\phi(u)du+\epsilon)^2}{(\int\phi(u)du)^2(\int\phi(u)du+\epsilon)^2}\right]\\
		&= \lim\limits_{\epsilon \to 0}\frac{1}{\epsilon}\left[\frac{-\epsilon\delta_x}{\int\phi(u)du+\epsilon}+\frac{\epsilon^2\phi(x)+2\epsilon\phi(x)\int\phi(u)du-\epsilon\delta_x(\phi(u)du)^2}{(\int\phi(u)du)^2(\int\phi(u)du+\epsilon)^2}\right]\\
		&= \lim\limits_{\epsilon \to 0}\left[-\frac{\delta_x}{\int\phi(u)du+\epsilon}+\frac{\epsilon\phi(x)+2\phi(x)\int\phi(u)du-\delta_x(\phi(u)du)^2}{(\int\phi(u)du)^2(\int\phi(u)du+\epsilon)^2}\right]\\
		&= \left[-\frac{\delta_x}{\int\phi(u)du}+\frac{2\phi(x)\int\phi(u)du-\delta_x(\phi(u)du)^2}{(\int\phi(u)du)^4}\right]\\
		&= \frac{2}{(\int\phi(u)du)^2}\left[p(x)-\delta_x\right]\label{for_freq_second_derivative}
\end{align}\end{linenomath*}
We can now substitute Eq. \ref{for_freq_first_derivative} and Eq. \ref{for_freq_second_derivative} into the RHS of Eq. \ref{density_Ito} to find an equation for the evolution of the trait frequency field. Once again, I will do this step by step for clarity. First, we evaluate the $dt$ term.

For the first term multiplying $dt$ in Eq. \ref{density_Ito}, we use Eq. \ref{for_freq_first_derivative} to find

\begin{linenomath*}\begin{align}
		\int\limits_{\mathcal{T}}\left\{\phi(y)w(y|\phi)+\mu Q(y|\phi)\right\}\frac{\delta \mathcal{H}_x}{\delta \phi(y)}dy &= \int\left\{\phi(y)w(y|\phi)+\mu Q(y|\phi)\right\}\frac{1}{\int\phi(u)du}\left[\delta_x-p(x)\right]dy\\
		&= \frac{1}{\int\phi(u)du}\bigg{[}\phi(x)w(x|\phi)-p(x)\int\phi(y)w(y)dy\nonumber\\
		&\hphantom{\frac{1}{\int\phi(u)du}\bigg{[}}+\mu\left( Q(x|\phi)-p(x)\int Q(y|\phi)dy\right)\bigg{]}\\
		&=\bigg{[}p(x)w(x|\phi)-p(x)\int p(y)w(y)dy\nonumber\\
		&+\frac{\mu}{\int\phi(u)du}\left( Q(x|\phi)-p(x)\int Q(y|\phi)dy\right)\bigg{]}
\end{align}\end{linenomath*}
Using the definition of mean fitness $\overline{w}$ from Eq. \ref{defn_mean_value}, we see that the first term on the RHS of Eq. \ref{density_Ito} is given by
\begin{linenomath*}\begin{align}
		&\int\limits_{\mathcal{T}}\left\{\phi(y)w(y|\phi)+\mu Q(y|\phi)\right\}\frac{\delta \mathcal{H}_x}{\delta \phi(y)}dy\nonumber\\
		&=[w(x|\phi)-\overline{w}(\phi, t)]p(x,t) + \frac{\mu}{\int\phi(u)du}\left( Q(x|\phi)-p(x)\int Q(y|\phi)dy\right)\label{freq_field_first_term}
\end{align}\end{linenomath*}

For the second term (second line) on the RHS of Eq. \ref{density_Ito}, we substitute Eq. \ref{for_freq_second_derivative} to find
\begin{linenomath*}\begin{align}
		&\frac{1}{2K}\left[\int\limits_{\mathcal{T}}  \left[\phi(y,t)\tau(y|\phi)+\mu Q(y|\phi)\right]\frac{\delta^2 \mathcal{H}_x}{\delta\phi(y,t)^2}dy\right]\nonumber\\
		&=\frac{1}{2K}\left[\int\left[\phi(y,t)\tau(y|\phi)+\mu Q(y|\phi)\right]\frac{2}{(\int\phi(u)du)^2}\left[p(x)-\delta_x\right]dy\right]\\
		&= \frac{1}{K(\int\phi(u)du)^2}\bigg{[}-\phi(x,t)\tau(x|\phi)+p(x)\int\tau(y|\phi)\phi(y)dy\nonumber\\
		&\hphantom{=\bigg{[}\frac{1}{K(\int\phi(u)du)^2}}+\mu\left(-Q(x|\phi)+p(x)\int Q(y|\phi)dy\right)\bigg{]}\\
		&= \frac{1}{K(\int\phi(u)du)}\bigg{[}-p(x,t)\tau(x|\phi)+p(x)\int\tau(y|\phi)p(y)dy\nonumber\\
		&\hphantom{=\bigg{[}\frac{1}{K(\int\phi(u)du)^2}}+\frac{\mu}{\int\phi(u)du}\left(-Q(x|\phi)+p(x)\int Q(y|\phi)dy\right)\bigg{]}
\end{align}\end{linenomath*}
Now using the definition of mean turnover rate $\overline{\tau}$ from Eq. \ref{defn_mean_value}, we see that the second term (second line) on the RHS of Eq. \ref{density_Ito} is given by
\begin{linenomath*}\begin{align}
		&\frac{1}{2K}\left[\int\limits_{\mathcal{T}}  \left[\phi(y,t)\tau(y|\phi)+\mu Q(y|\phi)\right]\frac{\delta^2 \mathcal{H}_x}{\delta\phi(y,t)^2}dy\right]\nonumber\\
		&= -\frac{1}{K\int\phi(u)du}\left[\left[\tau(x|\phi)-\overline{\tau}(t)\right]p(x)+\frac{\mu}{\int\phi(u)du}\left(Q(x|\phi)-p(x)\int Q(y|\phi)dy\right)\right]\label{freq_field_second_term}
\end{align}\end{linenomath*}
All that remains is to calculate the spacetime white noise term in Eq. \ref{density_Ito}. We find
\begin{linenomath*}\begin{align}
		&\frac{1}{\sqrt{K}}\int\limits_{\mathcal{T}} \sqrt{\phi(y,t)\tau(y|\phi)+\mu Q(y|\phi)}\frac{\delta \mathcal{H}_x}{\delta \phi(y,t)}\dot{W}(y,t)dy\nonumber\\
		&=\frac{1}{\sqrt{K}}\int\sqrt{\phi(y,t)\tau(y|\phi)+\mu Q(y|\phi)}\frac{1}{\int\phi(u)du}\left[\delta_x-p(x)\right]\dot{W}(y,t)dy
\end{align}\end{linenomath*}
and thus we see that the spacetime white noise term is given by
\begin{linenomath*}
	\begin{equation}
		\resizebox{\textwidth}{!}{$\displaystyle
			\begin{split}
				&\frac{1}{\sqrt{K}}\int\limits_{\mathcal{T}} \sqrt{\phi(y,t)\tau(y|\phi)+\mu Q(y|\phi)}\frac{\delta \mathcal{H}_x}{\delta \phi(y,t)}\dot{W}(y,t)dy \\
				&= \frac{1}{\sqrt{K}\int\phi(u)du}\left[\sqrt{\phi(x,t)\tau(x|\phi)+\mu Q(x|\phi)}\dot{W}(x,t)-p(x)\int\sqrt{\phi(y,t)\tau(y|\phi)+\mu Q(y|\phi)}\dot{W}(y,t)dy\right]
			\end{split}
			\label{freq_field_third_term}
			$
		}
	\end{equation}
\end{linenomath*}

Thus, using Eq. \ref{freq_field_first_term}, Eq. \ref{freq_field_second_term}, and Eq. \ref{freq_field_third_term} in Eq.~\ref{density_Ito}, we see that the trait frequency field obeys the SPDE
\begin{linenomath*}\begin{equation}
		\label{freq_field_supp}
		\frac{\partial p}{\partial t}(x,t) = \underbrace{\vphantom{\frac{1}{\sqrt{K}N_K(t)}}\mathcal{S}_w[x|\phi]}_{\substack{\text{Classical}\\\text{selection-}\\\text{mutation}}} - \underbrace{\vphantom{\frac{1}{\sqrt{K}N_K(t)}}\frac{1}{KN_K(t)}\mathcal{S}_{\tau}[x|\phi]}_{\substack{\text{Noise-induced}\\\text{selection-}\\\text{mutation}}} + \underbrace{\frac{1}{\sqrt{K}N_K(t)}\dot{W}_{p}(x,t)}_{\substack{\text{Undirected stochastic}\\\text{fluctuations}}}
\end{equation}\end{linenomath*}
where $N_K = \int\phi(u)du$ is the scaled total population size,
\begin{linenomath*}\begin{equation}
		\mathcal{S}_f[x| \phi] = \underbrace{\vphantom{{\frac{\mu}{N_K(t)}\left( Q(x|\phi)-p(x)\int\limits_{\mathcal{T}} Q(y|\phi)dy\right)}}(f[x|\phi] - \overline{f}(t))p(x,t)}_{\substack{\text{Selection for higher}\\\text{values of $f$}}} + \underbrace{\frac{\mu}{N_K(t)}\left( Q(x|\phi)-p(x)\int\limits_{\mathcal{T}} Q(y|\phi)dy\right)}_{\text{Mutation biases/transmission biases}}
\end{equation}\end{linenomath*}
represents selection-mutation balance for  $f$, and
\begin{linenomath*}\begin{equation}
		\resizebox{\textwidth}{!}{$\displaystyle\dot{W}_{p}(x,t) \coloneqq \sqrt{\phi(x,t)\tau(x|\phi)+\mu Q(x|\phi)}\dot{W}(x,t)-p(x)\int\limits_{\mathcal{T}}\sqrt{\phi(y,t)\tau(y|\phi)+\mu Q(y|\phi)}\dot{W}(y,t)dy$}
\end{equation}\end{linenomath*}

Eq.~\ref{freq_field_supp} is precisely Eq.~\ref{freq_field} in the main text.

\section{Recovering Kimura's continuum-of-alleles model from the continuous replicator-mutator equation}\label{App_kimura}

In this section, I recover Kimura's continuum of alleles model from equation Eq. \ref{cts_replicator_mutator}, the continuous replicator-mutator equation\footnote{obtained as the infinite population ($K \to \infty$) limit of Eq.~\ref{freq_field_supp}}, when mutation is modelled via convolution with a non-negative function that integrates to unity. To see this, let $Q(y|\phi) = (m\ast\phi)(y,t)$, where $m:\mathbb{R} \to [0,\infty)$ is the `mutation kernel', which by definition is non-negative and normalized such that $\int\limits_{\mathbb{R}}m(x)dx = 1$. Here, $\ast$ denotes convolution, defined for any two real functions $F$ and $G$ as
\begin{linenomath*}\begin{equation}
		(F\ast G) (x) = \int\limits_{\mathbb{R}}F(x-y)G(y)dy
\end{equation}\end{linenomath*}

Let us further recall that I assume (in the main text) that the population density process scales such that in the infinite population limit, $N_K(\cdot) = \int_{\mathcal{T}}\phi(x,\cdot)dx = 1$. Thus, both $m$ and $\phi(\cdot,t)$ are $L^{1}(\mathbb{R})$ functions. We can now use the general result that for any two $L^1(\mathbb{R})$ functions $F$ and $G$, the convolution $F \ast G$ satisfies:
\begin{linenomath*}\begin{equation}
		\int\limits_{\mathbb{R}}(F \ast G) (x)dx = \left(\int\limits_{\mathbb{R}} F(x)dx\right)\left(\int\limits_{\mathbb{R}}G(x)dx\right)
\end{equation}\end{linenomath*}
This general result is an easy consequence of the Fubini-Tonnelli theorem.

Using this result, the rightmost integral in Eq. \ref{cts_replicator_mutator} is
\begin{linenomath*}\begin{equation}
		\label{convolution_integral}
		\int\limits_{\mathbb{R}} Q(y|\phi)dy = \int\limits_{\mathbb{R}}(m\ast\phi)(y)dy  = \int\limits_{\mathbb{R}}\phi(y,t)dy\int\limits_{\mathbb{R}} m(y)dy   
\end{equation}\end{linenomath*}
Now, by definition of $N_K$, we have $\int\phi(y,t)dy = N_K$. Further, since $m$ is a kernel, it satisfies $\int\limits_{\mathbb{R}} m(y)dy = 1$. Equation Eq. \ref{convolution_integral} therefore becomes $\int\limits_{\mathbb{R}} Q(y|\phi)dy = N_K(t)$. Substituting this in Eq. \ref{cts_replicator_mutator}, we have
\begin{linenomath*}\begin{equation*}
		\frac{\partial p}{\partial t}(x,t) = \left[w(x|\phi) - \overline{w}(t)\right]p(x,t)+\frac{\mu}{N_K(t)}\left[\int\limits_{\mathbb{R}}m(x-z)\phi(z,t)dz - p(x,t)N_K(t)\right]
\end{equation*}\end{linenomath*}
Substituting our definition $p(z,t) = \phi(z,t)/N_K(t)$ now yields
\begin{linenomath*}\begin{equation}
		\frac{\partial p}{\partial t}(x,t) = \left[w(x|\phi) - \overline{w}(t)\right]p(x,t)+\mu\left[\int\limits_{\mathbb{R}}m(x-z)p(z,t)dz - p(x,t)\right]\label{kimura_continuum_model}
\end{equation}\end{linenomath*}
which is Kimura's continuum of alleles model~\citep{kimura_stochastic_1965,crow_introduction_1970}.

\section{Deriving an equation for the mean value of any type-level quantity}\label{App_stoch_Price}

In this section, I use the equation for the trait frequency field Eq. \ref{freq_field_supp} to derive an SDE for how the statistical mean value of any type level quantity changes in the population. I use the shorthand $\int = \int_{\mathcal{T}}$ 

Let $f(x|\phi)$ be any type-level quantity with statistical mean $\overline{f}(t)$. I allow for the possibility of $f$ varying through time even when $\phi$ is constant, to allow for the possibility of phenotypic plasticity and eco-evolutionary feedbacks. Assuming derivatives and integrals commute, we can calculate the rate of change of the mean value as

\begin{linenomath*}\begin{equation}
		\frac{d \overline{f}}{dt} = \int \frac{d}{dt}f(x|\phi)p(x,t)dx = \int f(x|\phi)\frac{\partial p}{\partial t}dx + \int p(x,t)\frac{\partial f}{\partial t}dx\label{intermediate_1_for_stoch_price}
\end{equation}\end{linenomath*}

We can now substitute the equation for $\partial p/\partial t$ from equation Eq. \ref{freq_field_supp} to calculate the first term on the RHS of Eq. \ref{intermediate_1_for_stoch_price}. We can thus write
\begin{linenomath*}\begin{equation}
		\label{intermediate_2_for_stoch_price}
		\int f(x|\phi)\frac{\partial p}{\partial t}dx =  \int f(x|\phi)\left(\mathcal{S}_w[x|\phi] - \frac{1}{KN_K(t)}\mathcal{S}_{\tau}[x|\phi] + \frac{1}{\sqrt{K}N_K(t)}\dot{W}_{p}(x,t)\right)dx
\end{equation}\end{linenomath*}
Given any type-level quantity $g(x|\phi)$, the selection-mutation operator for $g$ obeys
\begin{linenomath*}\begin{align}
		\int f(x|\phi)\mathcal{S}_g[x|\phi]dx &= \int f(x|\phi)(g(x|\phi) - \overline{g}(t))p(x,t)dx \nonumber\\
		&\hphantom{=}+ \frac{\mu}{N_K(t)} \left(\int f(x|\phi)\left( Q(x|\phi)-p(x,t)\int Q(y|\phi)dy\right)dx\right)\\
		&= \int f(x|\phi)g(x|\phi)p(x,t)dx-\left(\int f(x|\phi)dx\right)\overline{g}(t) \nonumber\\
		&\hphantom{=}+ \frac{\mu}{N_K(t)}\left(\int f(x|\phi) Q(x|\phi)dx-\int f(x|\phi)p(x,t)dx\int Q(y|\phi)dy\right)\\
		&= \overline{fg}-\overline{f}\overline{g} + \frac{\mu}{N_K}\left(\int f(x|\phi) Q(x|\phi)dx-\overline{f}\int Q(y|\phi)dy\right)
\end{align}\end{linenomath*}
Now using the definition of covariance from Eq. \ref{defn_covariance}, we thus see that we have the relation
\begin{linenomath*}\begin{equation}
		\label{selection_mutation_integral_for_price}
		\int f(x|\phi)\mathcal{S}_g[x|\phi]dx = \textrm{Cov}(g,f) + \frac{\mu}{N_K}\left(\int f(x|\phi) Q(x|\phi)dx-\overline{f}\int Q(y|\phi)dy\right)
\end{equation}\end{linenomath*}
Using Eq. \ref{selection_mutation_integral_for_price} in Eq. \ref{intermediate_2_for_stoch_price} for $g=w$ and $g=\tau$, we obtain
\begin{linenomath*}\begin{equation}
		\label{intermediate_3_for_stoch_price}
		\resizebox{\textwidth}{!}{$\displaystyle
			\begin{split}
				\int f(x|\phi)\frac{\partial p}{\partial t}dx &= \textrm{Cov}(w,f) - \frac{1}{KN_K}\textrm{Cov}(\tau,f)+\frac{\mu}{N_K}\left(1-\frac{1}{KN_K}\right)\left(\int f(x|\phi) Q(x|\phi)dx-\overline{f}\int Q(y|\phi)dy\right)\\
				&\hphantom{=}+\frac{1}{\sqrt{K}N_K(t)}\int f(x|\phi)\dot{W}_{p}(x,t)dx
			\end{split}
			$}
\end{equation}\end{linenomath*}
All that remains now is to calculate the stochastic integral term in Eq. \ref{intermediate_3_for_stoch_price}. Using Eq. \ref{freq_field_noise_term}, we find
\begin{linenomath*}\begin{align}
		&\int f(x|\phi)\dot{W}_{p}(x,t)dx\nonumber\\
		&= \int f(x|\phi)\left[\sqrt{\phi(x,t)\tau(x|\phi)+\mu Q(y|\phi)}\dot{W}(x,t)-p(x)\int\sqrt{\phi(y,t)\tau(y|\phi)+\mu Q(y|\phi)}\dot{W}(y,t)dy\right]dx\\
		&= \int f(x|\phi)\sqrt{\phi(x,t)\tau(x|\phi)+\mu Q(y|\phi)}\dot{W}(x,t)dx - \overline{f}(t)\int\sqrt{\phi(y,t)\tau(y|\phi)+\mu Q(y|\phi)}\dot{W}(y,t)dy\label{intermediate_3.5_for_stoch_price}
\end{align}\end{linenomath*}
Since $x$ and $y$ are both dummy variables that are being integrated over the entire space, we can therefore equivalently write Eq. \ref{intermediate_3.5_for_stoch_price} as
\begin{linenomath*}\begin{equation}
		\label{intermediate_4_for_stoch_price}
		\int f(x|\phi)\dot{W}_{p}(x,t)dx = \int \left(f(x|\phi)-\overline{f}(t)\right)\sqrt{\phi(x,t)\tau(x|\phi)+\mu Q(y|\phi)}\dot{W}(x,t)dx
\end{equation}\end{linenomath*}
Note that the spacetime white noise in Eq. \ref{intermediate_4_for_stoch_price} is being integrated over the entire domain of the space variable, and thus the resultant stochastic term is a stochastic integral with respect to a Brownian motion. In other words, the mean value $\overline{f}$ obeys a one-dimensional SDE. Defining
\begin{linenomath*}\begin{align}
		M_{\overline{f}}(p,N_K) &= \frac{\mu}{N_K}\left(1-\frac{1}{KN_K}\right)\left(\int f(x|\phi) Q(x|\phi)dx-\overline{f}\int Q(y|\phi)dy\right)\\
		\frac{dW_{\overline{f}}}{dt} &= \int \left(f(x|\phi)-\overline{f}(t)\right)\sqrt{\phi(x,t)\tau(x|\phi)+\mu Q(y|\phi)}\dot{W}(x,t)dx
\end{align}\end{linenomath*}
and substituting Eq. \ref{intermediate_3_for_stoch_price} into Eq. \ref{intermediate_1_for_stoch_price}, we thus obtain
\begin{linenomath*}\begin{equation}
		\label{stoch_price_supp_version}
		\frac{d \overline{f}}{dt} = \textrm{Cov}(w,f) - \frac{1}{KN_K}\textrm{Cov}(\tau,f) + M_{\overline{f}}(p,N_K) + \overline{ \left(\frac{\partial f}{\partial t}\right) } + \frac{1}{\sqrt{K}N_K}\frac{dW_{\overline{f}}}{dt}
\end{equation}\end{linenomath*}
which is equation Eq. \ref{stoch_Price} in the main text.

\section{A stochastic version of Fisher's fundamental theorem}\label{App_Fisher}

In this section, I derive a stochastic version of Fisher's fundamental theorem for finite, fluctuating populations and show that it indeed recovers Fisher's fundamental theorem in the infinite population limit. Let us first substitute $f(x|\phi) = w(x|\phi)$ and $\mu = 0$ into Eq. \ref{stoch_price_supp_version}. Note that since $\mu = 0$, the mutation term $M_{\overline{w}}$ vanishes. Thus, Eq. \ref{stoch_price_supp_version} becomes
\begin{linenomath*}\begin{equation}
		\label{intermediate_for_Fisher}
		\frac{d\overline{w}}{dt} = \sigma^2_w - \frac{1}{KN_K(t)}\textrm{Cov}(\tau,w) +  \overline{\left(\frac{\partial w}{\partial t}\right)}
		+ {\frac{1}{\sqrt{K}N_K(t)}\frac{dW_{\overline{w}}}{dt}}
\end{equation}\end{linenomath*}
where
\begin{linenomath*}\begin{equation}
		\frac{dW_{\overline{w}}}{dt} = \int\limits_{\mathcal{T}} \left(w(x|\phi)-\overline{w}(t)\right)\sqrt{\phi(x,t)\tau(x|\phi)}\dot{W}(x,t)dx
\end{equation}\end{linenomath*}
Now, since the covariance operator is bilinear, we can write
\begin{linenomath*}
	\begin{align}
		\textrm{Cov}(\tau,w) &= \textrm{Cov}(b^{\textrm{(ind)}} + d^{\textrm{(ind)}},b^{\textrm{(ind)}}-d^{\textrm{(ind)}})\\
		&= \textrm{Cov}(b^{\textrm{(ind)}},b^{\textrm{(ind)}}) - \textrm{Cov}(b^{\textrm{(ind)}},d^{\textrm{(ind)}}) + \textrm{Cov}(d^{\textrm{(ind)}},b^{\textrm{(ind)}}) - \textrm{Cov}(d^{\textrm{(ind)}},d^{\textrm{(ind)}})\\
		&= \sigma^2_{b^{\textrm{(ind)}}} - \sigma^2_{d^{\textrm{(ind)}}}\label{intermediate_2_for_Fisher}
	\end{align}
\end{linenomath*}
Let us now take probabilistic expectations $\mathbb{E}[\cdot]$ over the underlying probability space in Eq. \ref{intermediate_for_Fisher}. Upon doing this, ${dW_{\overline{w}}}/{dt}$ vanishes (by definition of white noise) and we therefore obtain
\begin{linenomath*}
	\begin{equation}
		\label{stoch_fisher}
		\mathbb{E}\left[\frac{d\overline{w}}{dt}\right] = 
		\negmedspace {\underbrace{\mystrut{4ex} \ \vphantom{ \frac{\sigma^2_{b^{\textrm{(ind)}}} - \sigma^2_{d^{\textrm{(ind)}}}}{KN_K(t)} } \mathbb{E}\left[\sigma^2_{w}\right] \ }_{\substack{\text{Fisher's} \\ \text{fundamental} \\ \text{theorem}}}} \ - \ {\underbrace{\mystrut{4ex}\mathbb{E}\left[\frac{\sigma^2_{b^{\textrm{(ind)}}} - \sigma^2_{d^{\textrm{(ind)}}}}{KN_K(t)}\right]}_{\substack{\text{Noise-induced} \\ \text{selection}}}} \ + {\underbrace{\mystrut{4ex} \vphantom{ \frac{\sigma^2_{b^{\textrm{(ind)}}} - \sigma^2_{d^{\textrm{(ind)}}}}{KN_K(t)} } \mathbb{E}\left[\hphantom{a}\overline{\hphantom{a}\frac{\partial w}{\partial t}\hphantom{a}}\hphantom{a}\right]}_{\substack{\text{Eco-evolutionary} \\ \text{feedbacks to fitness}}}}
	\end{equation}
\end{linenomath*}
where we have used the relation Eq. \ref{intermediate_2_for_Fisher}. This equation is a version of Fisher's fundamental theorem for finite, fluctuating populations. The first term on the RHS is the standard version of Fisher's fundamental theorem, the last term is present whenever we have eco-evolutionary feedbacks to fitness~\citep{frank_fishers_1992,kokko_stagnation_2021}, and the middle term represents the effects of noise-induced selection. Note that in finite populations, mean fitness can change directionally even when there is no standing variation in fitness ($\sigma^2_w = 0$) as long as $\sigma^2_{b^{\textrm{(ind)}}} \neq \sigma^2_{d^{\textrm{(ind)}}}$. Equation \ref{stoch_fisher} has been independently derived for discrete traits in a recent preprint focused on life-history evolution~\citep{kuosmanen_turnover_2022}.

\subsection*{The infinite population limit}

In the infinite population limit ($K \to \infty$), the expectations in equation Eq. \ref{stoch_fisher} become superfluous (since the process is now deterministic), and we thus obtain
\begin{linenomath*}
	\begin{equation}
		\frac{d\overline{w}}{dt} = \sigma^2_{w} + \overline{\hphantom{a}\frac{\partial w}{\partial t}\hphantom{a}}
	\end{equation}
\end{linenomath*}
which is Fisher's fundamental theorem in the presence of ecological changes to fitness~\citep{frank_fishers_1992,kokko_stagnation_2021}.

\section{Deriving a stochastic version of gradient dynamics}\label{App_gradient_dynamics}

In this section, I derive an equation of gradient dynamics for finite, fluctuating equations. I begin by once more listing my assumptions. I assume
\begin{itemize}
	\item Rare mutations, \emph{i.e.} $\mu$ is infinitesimally small.
	\item Small mutational effects with `almost faithful' reproduction, meaning $Q(x|\phi)$ is infinitesimally small.
	\item Strong selection, meaning that variants with low relative fitness are immediately eliminated from the population and the population remains sharply peaked around a small number of trait values. Mathematically, this means that if we begin with a monomorphic population $\phi(x,0) = N_{K}(0)\delta_{y_0}$ for some constants $N_K(0) > 0$ and $y_0 \in \mathcal{T}$, the population remains sufficiently well clustered for some time $t>0$ that we can continue to approximate the distribution $\phi(x,t)$ as a Dirac Delta mass $N_{K}(t)\delta_{y(t)}$ that is moving across the trait space according to a trajectory dictated by a function $y(t)$ (to be found). Note that we are not assuming that the population is exactly given by a delta mass, only that it is sufficiently sharply peaked that it can be approximated as one (i.e. `almost' every individual has trait value $y$).
\end{itemize}
I will call these together the strong-selection-weak-mutation (SSWM) assumption.
Let $\int = \int_{\mathcal{T}}$ for notational convenience. Under SSWM, we wish to obtain an equation for $y(t)$. Note first that we have:
\begin{linenomath*}
	\begin{equation}
		p(x,t) = \frac{\phi(x,t)}{N_K(t)} = \frac{N_{K}(t)\delta_{y(t)}}{N_K(t)} = \delta_y(t)
	\end{equation}
\end{linenomath*}
Further, for any type level quantity $f(x|\phi)$, we have
\begin{linenomath*}
	\begin{equation}
		\overline{f}(t) = \int f(x|\phi) p(x,t)dx = \int f(x|N_K\delta_{y(t)}) \delta_{y(t)} dx = f(y(t)|N_K\delta_{y(t)})
	\end{equation}
\end{linenomath*}
We can now substitute $f(x|\phi) = x$ into our stochastic Price equation (Eq. \ref{stoch_price_supp_version}) and, by virtue of our SSWM assumptions, neglect all terms of the form $\mu Q(x|\phi)$. Upon doing this and using the expressions for $\overline{x}$ and $p(x,t)$ from above, we obtain
\begin{linenomath*}
	\begin{equation}
		\label{intermediate_1_for_grad}
		\frac{dy}{dt} = \textrm{Cov}(w,x)-\frac{1}{KN_K(t)}\textrm{Cov}(\tau,x)+\frac{1}{\sqrt{K}N_K}\frac{dW_y}{dt}
	\end{equation}
\end{linenomath*}
where 
\begin{linenomath*}
	\begin{equation}
		\frac{dW_y}{dt} = \int\limits_{\mathcal{T}}(x-y(t))\sqrt{\tau(x|N_K(t)\delta_y(t))}\dot{W}(x,t)dx
	\end{equation}
\end{linenomath*}
Let $f(x|\phi)$ be any type-level quantity. We can calculate the statistical covariance of $f$ with $x$ under SSWM as
\begin{linenomath*}
	\begin{align}
		\textrm{Cov}(f,x) &= \int(f(x|\phi)-\overline{f})(x-\overline{x})p(x,t)dx\\
		&= \int(f(x|N_K\delta_{y(t)})-f(y(t)|N_K\delta_{y(t)}))(x-y(t))p(x,t)dx\label{cov_for_grad}
	\end{align}
\end{linenomath*}
Now, since under SSWM the population at time $t$ is sharply peaked around $y(t)$, we only need to worry about the value of $f(x|N_K\delta_{y(t)})$ at trait values that are (infinitesimally) close to $y(t)$. We can thus Taylor expand $f$ as:
\begin{linenomath*}
	\begin{equation}
		\label{taylor_for_grad}
		f(x|N_K\delta_{y(t)}) = f(y(t)|N_K\delta_{y(t)}) + (x-y)\frac{\partial f}{\partial z}(z|N_K\delta_{y(t)})\bigg{|}_{z=y} + \cdots 
	\end{equation}
\end{linenomath*}
and neglect all higher order terms. Using the relation Eq. \ref{taylor_for_grad} in Eq. \ref{cov_for_grad}, we obtain
\begin{linenomath*}
	\begin{align}
		\textrm{Cov}(f,x) &= \int(x-y)^2\frac{\partial f}{\partial z}(z|N_K\delta_{y(t)})\bigg{|}_{z=y}p(x,t)dx\\
		&= \int(x-y)^2p(x,t)dx\frac{\partial f}{\partial z}(z|N_K\delta_{y(t)})\bigg{|}_{z=y}\\
		&= \int(x-\overline{x})^2p(x,t)dx\frac{\partial f}{\partial z}(z|N_K\delta_{y(t)})\bigg{|}_{z=y}\\
		\Rightarrow \textrm{Cov}(f,x) &= \sigma^2_{x}(t)\frac{\partial f}{\partial z}(z|N_K\delta_{y(t)})\bigg{|}_{z=y}\label{intermediate_2_for_grad}
	\end{align}
\end{linenomath*}
Finally, using Eq. \ref{intermediate_2_for_grad} in Eq. \ref{intermediate_1_for_grad} with $f=w$ and $f=\tau$ yields
\begin{linenomath*}
	\begin{equation}
		\label{stoch_grad_supp}
		\frac{dy}{dt} = \sigma^2_{x}(t)\frac{\partial}{\partial z}\left[w(z|N_K\delta_{y(t)}) - \frac{1}{KN_K(t)}\tau(z|N_K\delta_{y(t)})\right]\bigg{|}_{z=y(t)} +\frac{1}{\sqrt{K}N_K}\frac{dW_y}{dt}
	\end{equation}
\end{linenomath*}
which is equation \ref{stoch_gradient} in the main text. Note that strictly speaking, if $\phi(x,t) = N_K\delta_{y(t)}$ exactly, then $\sigma^2_x \equiv 0$. This just reflects our assumption that mutations are vanishingly rare and mutants are sampled from infinitesimally close to the resident value, \emph{i.e.} that the population evolves in very small mutational steps. Thus, for our equation to make sense, the limits ${\mu \to 0}$, ${Q(x|\phi) \to 0}$, and ${\phi(x,t) \to \delta_{y(t)}N_K(t)}$ must be taken simultaneously such that $\sigma^2_{x}(t) \not\to 0$, a standard assumption in such heuristic derivations of gradient dynamics~\citep{dieckmann_dynamical_1996,lehtonen_price_2018}. More detailed mathematical arguments are required to ensure that this limit `makes sense'. Such convergence has been proved rigorously for the standard gradient equation using sophisticated mathematical tools grounded in martingale theory~\citep{champagnat_unifying_2006}.

\section{Detecting phenotypic clustering and speciation using spectral methods}\label{sec_Fourier}
In this section, I present a general analytical technique to study phenotypic clustering using the Fourier transform. To do this in full generality, we first require an approximation known as the weak noise approximation.

\subsection{The weak noise approximation }

If stochasticity takes the form of weak fluctuations about a deterministic trajectory, we can carry out a functional analog of the `linear noise approximation' or `weak noise approximation'~\citep{gardiner_stochastic_2009} to obtain a linear functional Fokker-Planck equation for the population density field. In the next section, we will use this linear approximation to illustrate how phenotypic clustering can be detected.

Assume that $\psi(x,t)$ is the deterministic trajectory of the density field in the infinite population limit, obtained as the solution to Eq. \ref{deterministic_traj}. Consider a new process $\{\zeta(\cdot,s)\}_{s \geq 0}$ which measures the fluctuations of $\phi(x,t)$ from the deterministic trajectory $\psi(x,t)$. More precisely, let us introduce the new variables:
\begin{linenomath*}\begin{equation}
		\begin{aligned}
			\label{functional_weak_noise_new_vars}
			\zeta(x,s) &= \sqrt{K}(\phi(x,t) - \psi(x,t))\\
			s &= t\\
			\Tilde{P}(\zeta,s) &= \frac{1}{\sqrt{K}}P(\phi,t)
		\end{aligned}
\end{equation}\end{linenomath*}
where we have introduced  a new time variable $s=t$ just to makes the substitutions/calculations cleaner when using the chain rule. Note that the following relations hold:
\begin{linenomath*}\begin{align}
		\frac{\delta F[\zeta]}{\delta \phi(x)} &= \int\limits_{\mathcal{T}}\frac{\delta F[\zeta]}{\delta \zeta(y)}\frac{\delta \zeta(y)}{\delta \phi(x)}dy = \sqrt{K}\frac{\delta F[\zeta]}{\delta \zeta(x)}\ \label{functional_weak_noise_first_subs}\\
		\frac{\partial}{\partial s} &= \frac{\partial}{\partial t}\label{functional_weak_noise_second_subs}
\end{align}\end{linenomath*}
Furthermore, for any $\zeta$, we have:
\begin{linenomath*}\begin{align}
		\frac{\partial \Tilde{P}}{\partial t}(\zeta,s) &= \frac{\delta \Tilde{P}}{\delta \zeta}\frac{\partial \zeta}{\partial t} + \frac{\partial \Tilde{P}}{\partial s}\frac{\partial s}{\partial t}\nonumber\\
		&=\frac{\delta \Tilde{P}}{\delta \zeta}\left(-\sqrt{K}\frac{\partial \psi}{\partial t}\right) + \frac{\partial \Tilde{P}}{\partial s}\nonumber\\
		&= -\sqrt{K}\frac{\delta}{\delta \zeta}\{\mathcal{A}^{-}(x|\psi)\Tilde{P}(\zeta,s)\} + \frac{\partial \Tilde{P}}{\partial s}\label{functional_weak_noise^{(t)}hird_subs}
\end{align}\end{linenomath*}
Reformulating equation Eq. \ref{density_FPE} in terms of the new variables Eq. \ref{functional_weak_noise_new_vars} and using the relations Eq. \ref{functional_weak_noise_first_subs}, Eq. \ref{functional_weak_noise_second_subs} and Eq. \ref{functional_weak_noise^{(t)}hird_subs}, we obtain:
\begin{linenomath*}\begin{equation*}
		\begin{split}
			-\sqrt{K}\frac{\delta}{\delta \zeta(x)}\{\mathcal{A}^{-}(x|\psi)\Tilde{P}(\zeta,s)\} + \frac{\partial \Tilde{P}}{\partial s} = \int\limits_{\mathcal{T}}\left[-
			\left(\sqrt{K}\frac{\delta }{\delta \zeta(x)}\right)\{\mathcal{A}^{-}\left(x\bigg{|}\psi+\frac{\zeta}{\sqrt{K}}\right)\tilde{P}(\zeta,s)\}\right]dx \\
			+ \int\limits_{\mathcal{T}}\left[\frac{1}{2K}\left(K\frac{\delta^2}{\delta\zeta(x)^2}\right)\{\mathcal{A}^{+}\left(x\bigg{|}\psi+\frac{\zeta}{\sqrt{K}}\right)\Tilde{P}(\zeta,s)\}\right]dx    
		\end{split}
\end{equation*}\end{linenomath*}
and rearranging gives us:
\begin{linenomath*}\begin{align}
		\label{functional_weak_noise_mid_expansion}
		\begin{split}
			\frac{\partial \Tilde{P}}{\partial s} &= -\sqrt{K}\int\limits_{\mathcal{T}}\frac{\delta }{\delta \zeta(x)}\left\{\left(\mathcal{A}^{-}\left(x\bigg{|}\psi+\frac{\zeta}{\sqrt{K}}\right)-\mathcal{A}^{-}(x|\psi)\right)\Tilde{P}(\zeta,s)\right\}dx\\
			&+\frac{1}{2}\int\limits_{\mathcal{T}}\frac{\delta^2}{\delta\zeta(x)^2}\{\mathcal{A}^{+}\left(x\bigg{|}\psi+\frac{\zeta}{\sqrt{K}}\right)\Tilde{P}(\zeta,s)\}dx
		\end{split}
\end{align}\end{linenomath*}
We will now Taylor expand our functionals about $\psi$ (I assume that this is possible). Thus, we have the expansions:
\begin{linenomath*}\begin{align*}
		\mathcal{A}^{-}\left(x\bigg{|}\psi+\frac{\zeta}{\sqrt{K}}\right) &= \mathcal{A}^{-}\left(x|\psi\right) + \frac{1}{\sqrt{K}}\int\limits_{\mathcal{T}}\zeta(y)\frac{\delta}{\delta \psi(y)}\{\mathcal{A}^{-}(y|\psi)\}dy + \cdots\\
		\mathcal{A}^{+}\left(x\bigg{|}\psi+\frac{\zeta}{\sqrt{K}}\right) &= \mathcal{A}^{+}\left(x|\psi\right) + \frac{1}{\sqrt{K}}\int\limits_{\mathcal{T}}\zeta(y)\frac{\delta}{\delta \psi(y)}\{\mathcal{A}^{+}(y|\psi)\}dy + \cdots
\end{align*}\end{linenomath*}
I also assume that $\tilde{P}$ can be expanded as
\begin{linenomath*}\begin{align*}
		\Tilde{P} &= \sum\limits_{n=0}^{\infty}\Tilde{P}_n\left(\frac{1}{\sqrt{K}}\right)^n
\end{align*}\end{linenomath*}
substituting these expansions into equation Eq. \ref{functional_weak_noise_mid_expansion}, equating coefficients of powers of $1/K$, and truncating at the lowest order term, we have:
\begin{linenomath*}\begin{equation*}
		\resizebox{\textwidth}{!}{$\displaystyle
			\frac{\partial \Tilde{P}_{0}}{\partial s}(\zeta,s) = \int\limits_{\mathcal{T}}\left[-\frac{\delta}{\delta \zeta(x)}\left\{\int\limits_{\mathcal{T}}\zeta(y)\frac{\delta}{\delta \psi(y)}\{\mathcal{A}^{-}(y|\psi)\}dy\Tilde{P}_{0}(\zeta,s)\right\}+\frac{1}{2}\mathcal{A}^{+}(x|\psi)\frac{\delta^2}{\delta\zeta(x)^2}\{\Tilde{P}_{0}(\zeta,s)\}\right]dx
			$}
\end{equation*}\end{linenomath*}
We thus arrive at the functional Fokker-Planck equation:
\begin{linenomath*}\begin{equation}
		\label{functional_WNE_zeroth_order}
		\frac{\partial \Tilde{P}_{0}}{\partial s}(\zeta,s) = \int\limits_{\mathcal{T}}\left(-\frac{\delta}{\delta \zeta(x)}\left\{\mathcal{D}_{\zeta}[\mathcal{A}^{-}](x)\Tilde{P}_{0}(\zeta,s)\right\}+\frac{1}{2}\mathcal{A}^{+}(x|\psi)\frac{\delta^2}{\delta\zeta(x)^2}\{\Tilde{P}_{0}(\zeta,s)\}\right)dx
\end{equation}\end{linenomath*}
where 
\begin{linenomath*}\begin{equation*}
		\mathcal{D}_{\zeta}[\mathcal{A}^{-}](x) = \int\limits_{\mathcal{T}}\zeta(y)\frac{\delta}{\delta \psi(y)}\{\mathcal{A}^{-}(y|\psi)\}dy = \frac{d}{d\epsilon}\mathcal{A}^-(x|\psi + \epsilon \zeta) \bigg{|}_{\epsilon = 0}
\end{equation*}\end{linenomath*}
can be thought of as the functional analog of a directional derivative of $\mathcal{A}^-(x|\psi)$ `in the direction of the field' $\zeta$. We will now use this linear approximation to study phenotypic clustering/evolutionary branching using spectral methods first introduced for various specific models by~\citet{rogers_demographic_2012} and further expanded upon for some specific models of adaptive dynamics in~\citet{rogers_modes_2015}.

\subsection{Detecting phenotypic clusters through Fourier analysis}

We begin with the linear functional Fokker-Planck equation
\begin{linenomath*}\begin{equation}
		\label{functional_WNE_for_App}
		\frac{\partial P}{\partial t}(\zeta,t) = \int\limits_{\mathcal{T}}\left(-\frac{\delta}{\delta \zeta(x)}\left\{\mathcal{D}_{\zeta}[\mathcal{A}^{-}](x)P(\zeta,t)\right\}+\frac{1}{2}\mathcal{A}^{+}(x|\psi)\frac{\delta^2}{\delta\zeta(x)^2}\{P(\zeta,t)\}\right)dx
\end{equation}\end{linenomath*}
for describing stochastic fluctuations $\zeta$ from the deterministic solution obtained by solving Eq. \ref{deterministic_traj}. Our goal is now to find a method to effectively detect and describe evolutionary branches (modes in trait space, corresponding to individual morphs) for this process. We will do this by measuring the autocorrelation of the field $\phi$, a task made easier by moving to Fourier space.

\myfig{0.8}{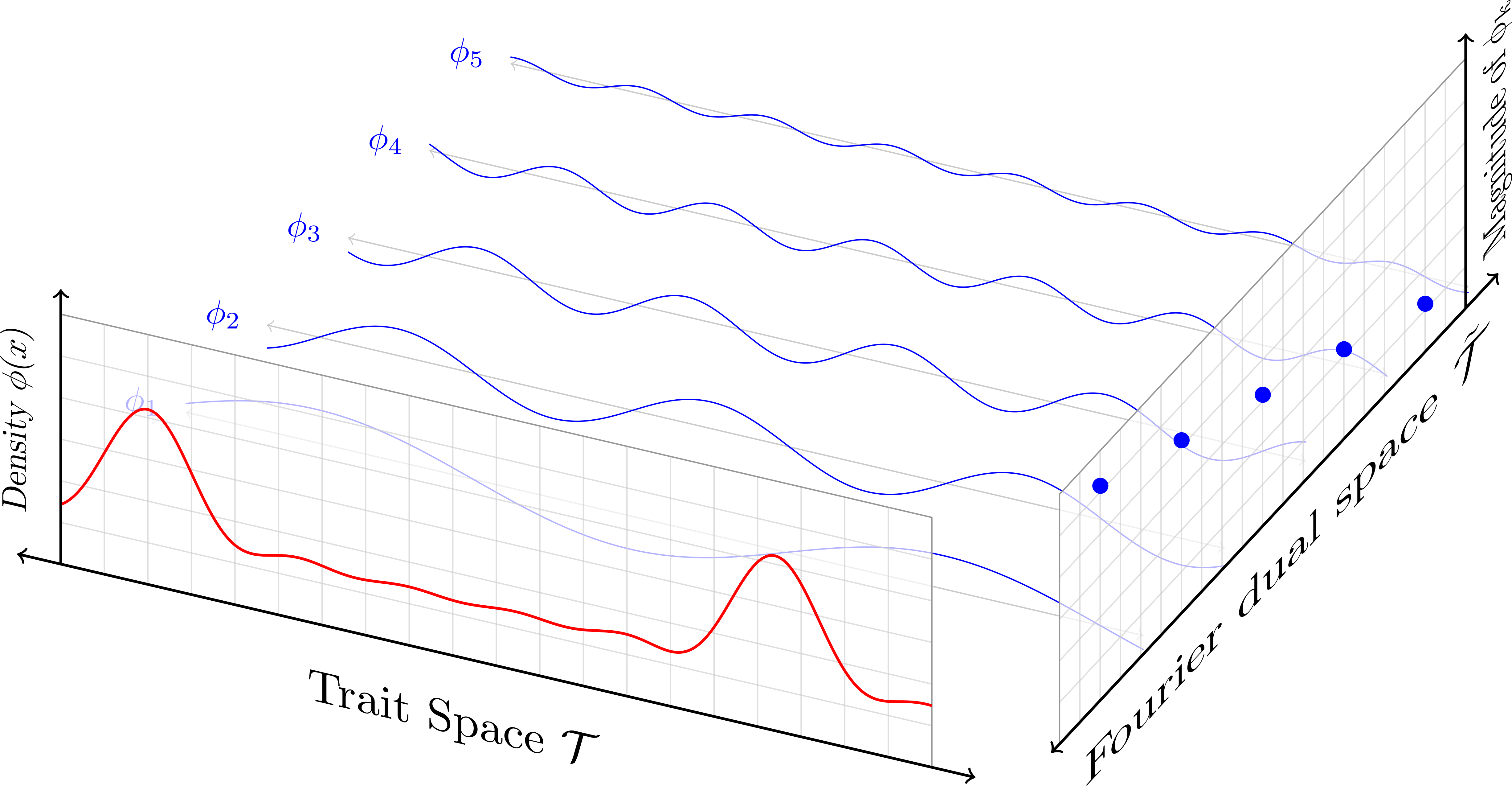}{\textbf{Schematic description of Fourier decomposition}. Any density field $\color{red}\phi$ (shown in red) can be decomposed into a sum of infinitely many Fourier modes $\color{blue}\phi_k$ (shown in blue). In the Fourier dual space, we can look at the peaks of each of these Fourier modes: The magnitude of $\color{blue}\phi_k$ tells us how much it contributes to the actual function of interest $\color{red}\phi$. The autocorrelation of $\color{red}\phi$ in the trait space can be measured using the power spectral density in the Fourier dual space.}{fig_Fourier}
A convenient theorem due to Weiner and Khinchin relates the autocorrelation of a probability distribution to its power spectral density via Fourier transformation. I will thus restrict myself to cases in which we can express our focal function $\phi$ in terms of the Fourier basis $\{e^{ikx}\}_{k\in\mathbb{Z}}$ (Figure \ref{fig_Fourier}). For example, if $\mathcal{T}$ is an interval, this can be done by imposing `periodic boundary conditions' (\emph{i.e.}extending all functions from $\mathcal{T}$ to $\mathbb{R}$ in a way that they appear periodic with period given by the length of the interval $\mathcal{T}$). 
If $\mathcal{D}_{\zeta}[\mathcal{A}^{-}]$ is a translation-invariant linear operator, then $\exp(ikx)$ acts as an eigenfunction, significantly simplifying the calculations. I therefore assume that $\mathcal{D}_{\zeta}[\mathcal{A}^{-}]$ takes the form:
\begin{linenomath*}\begin{equation*}
		\mathcal{D}_{\zeta}[\mathcal{A}^-](x,t) = L[\zeta(x,t)]   
\end{equation*}\end{linenomath*}
for a translation-invariant linear operator $L$ that only depends on $x$ and $t$. This is not as restrictive as it initially sounds. For example, both the Laplacian operator and the convolution operator are linear and translation invariant. The presence of phenotypic clustering and polymorphisms can be analyzed by examining the power spectrum of $P(\zeta,t)$ over the trait space, which is precisely what we will do.

Let the Fourier basis representations $\zeta$, and $\mathcal{A}^{+}(x|\psi)$ be given by:
\begin{linenomath*}\begin{equation}
		\label{fourier_representations_functions}
		\begin{aligned}
			\zeta(x,t) &= \sum\limits_{k=-\infty}^{\infty}e^{ikx}\zeta_k(t) \ \ ; \ \ \zeta_k(t) = \int\limits_{\mathcal{T}}\zeta(x,t)e^{-ikx}dx\\
			\mathcal{A}^{+}(x|\psi) &= \sum\limits_{k=-\infty}^{\infty}e^{ikx}A_k(t) \ \ ; \ \ A_k(t) = \int\limits_{\mathcal{T}}\mathcal{A}^{+}(x|\psi)e^{-ikx}dx
		\end{aligned}
\end{equation}\end{linenomath*}
In this case, the functional derivative operator obeys:
\begin{linenomath*}\begin{equation}
		\label{fourier_representations_derivative}
		\frac{\delta}{\delta \zeta(x)} = \sum\limits_{k=-\infty}^{\infty}e^{-ikx}\frac{\partial}{\partial \zeta_k}
\end{equation}\end{linenomath*}
and since $L$ is linear and translation-invariant, we also have the relation\footnote{This is because $\exp(ikx)$ acts as an eigenfunction for translation invariant linear operators, and therefore, for any function $\varphi = \sum\varphi_k\exp(ikx)$, we have the relation $L[\varphi] = L[\sum\varphi_k\exp(ikx)]=\sum\varphi_kL[\exp(ikx)]=\sum\varphi_kL_k\exp(ikx)$, where $L_k$ is the eigenvalue of $L$ associated with the eigenfunction $\exp(ikx)$. It is helpful to draw the analogy with eigenvectors of matrices and view $L_k\varphi_k$ as the projection of $L[\varphi]$ along the $k$th eigenvector $e_k = \exp(ikx)$.}:
\begin{linenomath*}\begin{equation}
		\label{fourier_representation_linear_operator}
		L[\zeta] = \sum\limits_{k=-\infty}^{\infty}L_{k}\zeta_ke^{ikx}
\end{equation}\end{linenomath*}
where 
\begin{linenomath*}\begin{equation*}
		L_k = e^{-ikx}L[e^{ikx}]
\end{equation*}\end{linenomath*}
Lastly, by definition of Fourier modes, we have, for any differentiable real function $F$ and any fixed time $t > 0$:
\begin{linenomath*}\begin{equation}
		\label{fourier_mode_relation}
		\frac{\partial}{\partial \zeta_j(t)}F(\zeta_i(t)) = \delta_{i,j}F'(\zeta_j(t))
\end{equation}\end{linenomath*}
where $\delta_{i,j}$ is the Kronecker delta, defined as
\begin{equation*}
	\delta_{i,j} = \begin{cases}
		1 & i=j\\
		0 & i\neq j
	\end{cases}
\end{equation*}
Using Eq. \ref{fourier_representations_functions}, Eq. \ref{fourier_representations_derivative}, and Eq. \ref{fourier_representation_linear_operator} in Eq. \ref{functional_WNE_for_App}, we get, for the first term of the RHS:
\begin{gather}
	-\int\limits_{\mathcal{T}}\frac{\delta}{\delta \zeta(x)}\left\{L[\zeta(x,t)]P(\zeta,t)\right\}dx\nonumber\\
	= -\int\limits_{\mathcal{T}}\sum\limits_{k}e^{-ikx}\frac{\partial}{\partial \zeta_k}\{\sum\limits_{n}e^{inx}L_n\zeta_nP\}dx\nonumber\\
	= -\int\limits_{\mathcal{T}}\sum\limits_{k}\sum\limits_{n}e^{-i(k-n)x}\frac{\partial}{\partial \zeta_k}\{L_n\zeta_nP\}dx\nonumber\\
	= -2\pi\sum\limits_{k}L_{k}\frac{\partial}{\partial \zeta_k}\{\zeta_kP\}\label{fourier_FPE_first_term}
\end{gather}
and for the second:
\begin{gather}
	\int\limits_{\mathcal{T}}\sum\limits_{k}e^{ikx}A_k\left(\sum\limits_{m}\sum\limits_{n}e^{-i(m+n)x}\frac{\partial}{\partial \zeta_m}\frac{\partial}{\partial \zeta_n}P\right)dx\nonumber\\
	= \int\limits_{\mathcal{T}}\sum\limits_{k}\sum\limits_{m}\sum\limits_{n}e^{i(k-m-n)x}A_k\frac{\partial}{\partial \zeta_m}\frac{\partial}{\partial \zeta_n}\{P\}dx\nonumber\\
	= 2\pi\sum\limits_{m}\sum\limits_{n}A_{m+n}\frac{\partial}{\partial \zeta_m}\frac{\partial}{\partial \zeta_{n}}\{P\}\label{fourier_FPE_second_term}
\end{gather}
Substituting Eq. \ref{fourier_FPE_first_term} and Eq. \ref{fourier_FPE_second_term} into Eq. \ref{functional_WNE_for_App}, we see that the Fokker-Planck equation Eq. \ref{functional_WNE_for_App} in Fourier space reads:
\begin{linenomath*}\begin{equation}
		\label{fourier_FPE}
		\frac{\partial P}{\partial t} = -2\pi\sum\limits_{k}L_{k}\frac{\partial}{\partial \zeta_k}\{\zeta_kP\} + \pi\sum\limits_{m}\sum\limits_{n}A_{m+n}\frac{\partial}{\partial \zeta_m}\frac{\partial}{\partial \zeta_{n}}\{P\}
\end{equation}\end{linenomath*}
Multiplying both sides of Eq. \ref{fourier_FPE} by $\zeta_r$ and integrating over the probability space to obtain expectation values, we see that
\begin{linenomath*}\begin{align}
		\frac{d}{dt}\mathbb{E}[\zeta_r] &= - 2\pi \sum\limits_{k}\int\zeta_rL_k\frac{\partial}{\partial \zeta_k}\{\zeta_k P\}d\omega + \pi\sum\limits_{m}\sum\limits_{n}A_{m+n}\int\zeta_r\frac{\partial}{\partial \zeta_m}\frac{\partial}{\partial \zeta_{n}}(P)d\omega\label{intermediate_in_fourier_for_int_by_parts}
\end{align}\end{linenomath*}
We will evaluate the terms on the RHS of Eq. \ref{intermediate_in_fourier_for_int_by_parts} using integration by parts. Recall that for any two functions $u$ and $v$ defined on a domain $\Omega$, the general formula for integration by parts is given by:
\begin{linenomath*}\begin{equation}
		\label{int_by_parts_general_formula}
		\int\limits_{\Omega}\frac{\partial u}{\partial x_i}vd\mathbf{x} = -\int\limits_{\Omega}u\frac{\partial v}{\partial x_i}d\mathbf{x} + \int\limits_{\partial\Omega}uv\gamma_{i}dS(\mathbf{x})
\end{equation}\end{linenomath*}
where $\partial \Omega$ is the boundary of $\Omega$, $dS$ is the surface element of this boundary, and $\gamma_i$ is the $i\textsuperscript{th}$ component of the unit outward normal to the boundary. In our case, I assume that the probability of extreme events is negligible enough that $P$ decays rapidly near the boundaries and we can neglect the contributions of the boundary term (second term on the RHS of Eq. \ref{int_by_parts_general_formula}). Thus, using integration by parts and neglecting the boundary terms on the RHS of equation Eq. \ref{intermediate_in_fourier_for_int_by_parts}, we obtain
\begin{linenomath*}\begin{align}
		\frac{d}{dt}\mathbb{E}[\zeta_r]	&=  2\pi \sum\limits_{k}L_k\int\zeta_k\frac{\partial \zeta_r}{\partial \zeta_k}Pd\omega + \pi\sum\limits_{m}\sum\limits_{n}A_{m+n}\int\frac{\partial^2 \zeta_r}{\partial \zeta_m\partial \zeta_{n}}Pd\omega =  2\pi L_{r}\mathbb{E}[\zeta_r]\label{fourier_mode_mean}
\end{align}\end{linenomath*}
where we have used the relation in Eq. \ref{fourier_mode_relation} to arrive at the final expression. Similarly, multiplying Eq. \ref{fourier_FPE} by $\zeta_r\zeta_s$, integrating over the probability space and using integration by parts, we get:
\begin{linenomath*}\begin{align}
		\frac{d}{dt}\mathbb{E}[\zeta_r\zeta_s] &= 2\pi \sum\limits_{k}L_{k}\int\zeta_kP\frac{\partial}{\partial \zeta_k}\{\zeta_r\zeta_s\}d\omega + \pi\sum\limits_{m}\sum\limits_{n}A_{m+n}\int P\frac{\partial}{\partial \zeta_m}\frac{\partial}{\partial \zeta_{n}}\{\zeta_r\zeta_s\}d\omega\nonumber\\
		&= 2\pi (L_{r} + L_{s})\mathbb{E}[\zeta_r\zeta_s] + \pi (A_{2r}+A_{2s})\label{fourier_mode_covariance}
\end{align}\end{linenomath*}
At the stationary state, the LHS must be zero by definition, and we must therefore have, for every $r,s \in \mathbb{Z}$,:
\begin{linenomath*}\begin{equation}
		\label{fourier_mode_covariance_stationary}
		\mathbb{E}[\zeta_r\zeta_s] = -   \frac{A_{2r}+A_{2s}}{2(L_{r}+L_{s})}
\end{equation}\end{linenomath*}
Recall now that the Fourier modes of any real function $\varphi$ must satisfy $\varphi_{-r} = \overline{\varphi}_r$. Since $\zeta$, $A$ and $L$ are all real, we can substitute $s=-r$ in equation Eq. \ref{fourier_mode_covariance_stationary} to obtain the autocovariance relation:
\begin{linenomath*}\begin{equation}
		\label{fourier_mode_autocovariance}
		\mathbb{E}[|\zeta_r|^2] =- \frac{\mathrm{Re}(A_{2r})}{2\mathrm{Re}(L_{r})}
\end{equation}\end{linenomath*}

The presence of phenotypic clustering can be detected using the `spatial covariance' of our original process $\phi$, defined as~\citep{rogers_demographic_2012}:
\begin{linenomath*}\begin{equation}
		\label{spatial_covariance_defn}
		\Xi[x] = m(\mathcal{T})\int\limits_{\mathcal{T}}\mathbb{E}[\phi_{\infty}(x)\phi_{\infty}(y-x)]dy
\end{equation}\end{linenomath*}
where $\phi_{\infty}$ is the stationary state distribution of $\{\phi_t\}_{t}$ and $m$ is the Lebesgue measure. We can use a spatial analogue of the Wiener-Khinchin theorem to calculate~\citep{rogers_demographic_2012,rogers_modes_2015}:
\begin{linenomath*}\begin{equation}
		\label{spatial_covariance_zeta}
		\Xi[x] = m(\mathcal{T})\bigg[\underbrace{\int\limits_{\mathcal{T}}\psi_{\infty}(x)\psi_{\infty}(y-x)dy}_{\text{Infinite population prediction}} + \underbrace{\vphantom{\int\limits_{\mathcal{T}}}\frac{1}{K}\sum\limits_{r=-\infty}^{\infty}\mathbb{E}[|\zeta_r|^2]e^{irx}}_{\text{Finite population corrections}}\bigg]
\end{equation}\end{linenomath*}
where the expectations in the second term are for the stationary state. A flat $\Xi[x]$ indicates that there are no clusters, and peaks indicate the presence of clusters. Notice that we generically expect stochastic fluctuations to induce some finite population corrections to the spatial covariance.

\section{An example: Asexual resource competition}\label{App_example}

In this section, I present an example of a model of resource competition to illustrate the analytical pipeline outlined in this paper. Let us imagine a population of asexual individuals bearing some quantitative trait $x$ taking values in $\mathbb{R}$. I assume the particular birth and death rate functionals
\begin{linenomath*}
	\begin{equation}
		\label{example_b_d_functions}
		\begin{aligned}
			b(x|\nu) &= \int M(x,y)\nu(y,t) dy \ \ ; \ \ M(x,y) = \exp\left(\frac{-(x-y)^2}{\sigma^2_m}\right)\\
			d(x|\nu) &= \frac{\nu(x,t)}{K} \int \alpha(x,y)\nu(y,t) dy \ \ ; \ \ \alpha(x,y) = \exp\left(\frac{-(x-y)^2}{\sigma^2_{\alpha}}\right)
		\end{aligned}
	\end{equation}
\end{linenomath*}

where the integrals are over the entire real line.

The interpretation is as follows: Individuals give birth to offspring with mutations. The effect of the mutations is parametrized by the `mutation kernel' $M$. $M(x,y)$ represents the per-capita rate at which type $y$ individuals give birth to type $x$ individuals. I assume $M$ has the functional form of a Normal distribution centered around the focal trait value and with a variance of $\sigma^2_m$. Thus, birth is most often with no mutation, and more extreme mutations are less likely. The death rate functional incorporates the effect of death due to resource competition. The effects of competition are parametrized by a competition kernel $\alpha$. The quantity $\alpha(x,y)$ represents the per-capita additional death rate of type $x$ individuals due to competition with type $y$ individuals. I have assumed that the strength of competition effects takes the form of a Normal distribution about the focal trait value, with variance $\sigma^2_{\alpha} > 0$. Thus, individuals experience less competition from individuals who are at a greater phenotypic distance away from them in trait space. The death rate term also incorporates a carrying capacity $K > 0$. This functional form for the death rate can be seen as the continuous analog of the death rate term that arises in Lotka-Volterra competition ($x_i\sum_j \alpha_{ij}x_j/K$).

Switching from population numbers $\nu$ to the population density field $\phi = \nu/K$ via the variable transformation presented in Eq. \ref{pop_density_BD_variable_transform}, we obtain the new stochastic field equations paramterized by the scaled birth and death rate functionals:
\begin{linenomath*}
	\begin{equation}
		\label{example_b_d_functions_density_exact}
		\begin{aligned}
			b(x|\phi) &= \int M(x,y)\phi(y,t) dy\\
			d(x|\phi) &= \phi(x,t) \int \alpha(x,y)\phi(y,t) dy
		\end{aligned}
	\end{equation}
\end{linenomath*}
Assuming the mutational variance $\sigma^2_m$ is small, we can approximate the birth rate functional as
\begin{linenomath*}
	\begin{equation}
		\int e^{-(x-y)^2/{\sigma^2_m}}\phi(y,t)dy = \phi(x,t) + \frac{\sigma^2_m}{2}\frac{\partial^2\phi}{\partial x^2} + \ldots
	\end{equation}
\end{linenomath*}
Discarding higher order terms, we can thus write the birth and death rates presented in Eq. \ref{example_b_d_functions_density_exact} as
\begin{linenomath*}
	\begin{equation}
		\label{example_b_d_functions_density}
		\begin{aligned}
			b(x|\phi) &= \phi(x,t) + \frac{\sigma^2_m}{2}\nabla^2_x\phi \\
			d(x|\phi) &= \phi(x,t) \int \alpha(x,y)\phi(y,t) dy 
		\end{aligned}
	\end{equation}
\end{linenomath*}
where we have used the notation $\nabla^2_x = \partial^2/\partial x^2$ for convenience. We can now compare terms with the definitions of fitness, turnover, and mutational effects to identify:
\begin{linenomath*}
	\begin{align}
		w(x|\phi) &= 1-\int \alpha(x,y)\phi(y,t) dy \label{example_fitness}\\
		\tau(x|\phi) &= 1+\int \alpha(x,y)\phi(y,t) dy \label{example_turnover}\\
		\mu &= \sigma^2_m/2\label{example_mutation_rate}\\
		Q(x|\phi) &= \nabla^2_x\phi \label{example_mutational_term}
	\end{align}
\end{linenomath*}
These quantities are sufficient to derive all the general equations presented in this paper. For conciseness, I only present the resultant equations for the density field (Eq. \ref{density_SPDE}) and the stochastic gradient equation (Eq. \ref{stoch_gradient}) below.

\subsection{The population density field}

From Eq. \ref{density_SPDE}, we see that the population density field obeys the SPDE
\begin{linenomath*}
	\begin{equation}
		\label{example_density_field}
		\begin{split}
			\frac{\partial \phi}{\partial t}(x,t) &= \phi(x,t)\left(1-\int \alpha(x,y)\phi(y,t) dy\right) + \frac{\sigma^2_m}{2}\nabla^2_x\phi\\
			\hphantom{=}&+ \frac{1}{\sqrt{K}}\sqrt{\phi(x,t)\left(1+\int \alpha(x,y)\phi(y,t) dy\right) + \frac{\sigma^2_m}{2}\nabla^2_x\phi\hphantom{+}}\dot{W}(x,t)
		\end{split}
	\end{equation}
\end{linenomath*}
Equation \ref{example_density_field} is a stochastic version of the quantitative logistic equation that includes the effects of noise-induced selection and drift. This can be seen by examining the infinite population limit.

\subsubsection*{The infinite population limit}
In the infinite population limit, we see from Eq. \ref{deterministic_traj} that the density field obeys the PDE
\begin{linenomath*}
	\begin{equation}
		\label{example_dens_inf_limit}
		\frac{\partial \phi}{\partial t}(x,t) = \phi(x,t)\left(1-\int \alpha(x,y)\phi(y,t) dy\right) + \frac{\sigma^2_m}{2}\nabla^2_x\phi
	\end{equation}
\end{linenomath*}
Equation \ref{example_dens_inf_limit} is the so-called `quantitative logistic equation' and has been used to model asexual resource competition~\citep{doebeli_adaptive_2011}.

\subsection{The stochastic gradient equation}

Let us first calculate the relevant fitness and turnover functions. From Eq. \ref{example_fitness}, we obtain:
\begin{linenomath*}
	\begin{equation}
		\label{example_fitness_SSWM}
		w(x|N_K\delta_y) = 1-\int \alpha(x,z)N_K\delta_y dz = 1-N_K\alpha(x,y)
	\end{equation}
\end{linenomath*}
and similarly, from Eq. \ref{example_turnover}, we obtain
\begin{linenomath*}
	\begin{equation}
		\label{example_turnover_SSWM}
		\tau(x|N_K\delta_y) = 1+\int \alpha(x,z)N_K\delta_y dz = 1+N_K\alpha(x,y)
	\end{equation}
\end{linenomath*}
Substituting Eq. \ref{example_fitness_SSWM} and Eq. \ref{example_turnover_SSWM} into Eq. \ref{stoch_grad_supp}, we thus obtain the stochastic gradient equation:
\begin{linenomath*}
	\begin{equation}
		\label{example_stoch_gradient}
		\frac{dy}{dt} = \sigma^2_x(t) \frac{\partial G(x;y)}{\partial x}\bigg{|}_{x=y} + \frac{dW_{y}}{dt}
	\end{equation}
\end{linenomath*}
where
\begin{linenomath*}
	\begin{equation}
		G(x;y) = 1 - \frac{1}{KN_K(t)} - \left(\frac{1+KN_K(t)}{K}\right)\alpha(x,y)
	\end{equation}
\end{linenomath*}
and
\begin{linenomath*}
	\begin{equation}
		\frac{dW_{y}}{dt}	= \int(x-y(t))\sqrt{1+N_K(t)\alpha(x,y)}\dot{W}(x,t)dx
	\end{equation}
\end{linenomath*}
Equation \ref{example_stoch_gradient} is a gradient equation that incorporates the effects of noise-induced selection and genetic drift, and thus can be used to study evolution in finite populations. The infinite population limit recovers a well-known model from adaptive dynamics, as I show below.

\subsubsection*{The infinite population limit}
In the infinite population limit, we see from Eq. \ref{stoch_grad_supp} and Eq. \ref{example_fitness_SSWM}  that we obtain the gradient equation
\begin{linenomath*}
	\begin{equation}
		\label{example_gradient_infpop}
		\frac{dy}{dt} = - \sigma^2_x(t) \frac{\partial \alpha(x,y)}{\partial x}\bigg{|}_{x=y}
	\end{equation}
\end{linenomath*}
Equation \ref{example_gradient_infpop} is the asexual model of sympatric speciation due to resource competition presented in Chapter 3 of~\cite{doebeli_adaptive_2011}.

\section{    A stochastic Fisher-KPP equation from an individual-based model}\label{App_FKPP}

Consider the birth and death rate functionals given by \ref{example_b_d_functions} with the competition kernel $\alpha(x,y) = \delta_x$. Thus, we can write
\begin{linenomath*}
	\begin{equation}
		\label{example_b_d_functions_FKPP}
		\begin{aligned}
			b(x|\nu) &= \int M(x,y)\nu(y,t) dy \ \ ; \ \ M(x,y) = \exp\left(\frac{-(x-y)^2}{\sigma^2_m}\right)\\
			d(x|\nu) &= \frac{\nu(x,t)}{K} \int \delta_x\nu(y,t) dy = \frac{1}{K}\nu^2(x,t)
		\end{aligned}
	\end{equation}
\end{linenomath*}

Biologically, $\alpha(x,y) = \delta_x$ means that individuals only compete with other individuals that have the exact same phenotype as them, and can be thought of as taking the limit of infinitely narrow niche width in the competition model introduced in the previous section. For instance, such dynamics could be a reasonable model if the trait in question is location in physical space. Using the same approximation for the birth rate as in the previous section, we arrive at the birth and death rates in density space

\begin{linenomath*}
	\begin{equation}
		\label{example_b_d_functions_density_FKPP}
		\begin{aligned}
			b(x|\phi) &= \phi(x,t) + \frac{\sigma^2_m}{2}\nabla^2_x\phi \\
			d(x|\phi) &= \phi^2(x,t)
		\end{aligned}
	\end{equation}
\end{linenomath*}

and thus

\begin{linenomath*}
	\begin{align}
		w(x|\phi) &= 1-\phi(x,t)\\
		\tau(x|\phi) &= 1+\phi(x,t)\\
		\mu &= \sigma^2_m/2\\
		Q(x|\phi) &= \nabla^2_x\phi
	\end{align}
\end{linenomath*}

We can now subtitute these functions into the population level equations for population densities (abundances).

\subsection{The population density field}

From Eq. \ref{density_SPDE}, we see that the population density field obeys the SPDE
\begin{linenomath*}
	\begin{equation}
		\label{example_stoch_Fisher_KPP}
		\begin{split}
			\frac{\partial \phi}{\partial t}(x,t) &= \phi\left(1-\phi\right) + \frac{\sigma^2_m}{2}\nabla^2_x\phi + \frac{1}{\sqrt{K}}\sqrt{\phi\left(1+\phi\right) + \frac{\sigma^2_m}{2}\nabla^2_x\phi\hphantom{+}}\dot{W}(x,t)
		\end{split}
	\end{equation}
\end{linenomath*}
Equation \ref{example_stoch_Fisher_KPP} is similar to a stochastic analog of the Fisher-KPP equation on $\mathbb{R}$, as will become clear upon taking the infinite population limit below. However, note that Eq.~\ref{example_stoch_Fisher_KPP} is \emph{not} the SPDE that is usually referred to as the `stochastic Fisher-KPP' equation~\citep{doering_interacting_2003,barton_modelling_2013}. The usual `stochastic Fisher-KPP' equation is instead the SPDE~\citep{doering_interacting_2003,barton_modelling_2013}

\begin{linenomath*}
	\begin{equation}
		\label{classical_stoch_Fisher_KPP}
		\begin{split}
			\frac{\partial u}{\partial t}(x,t) &= u\left(1-u\right) + \frac{\sigma^2_m}{2}\nabla^2_xu + \epsilon \sqrt{u(1-u)}\dot{W}(x,t)
		\end{split}
	\end{equation}
\end{linenomath*}

The discrepancy arises because the SPDE \eqref{classical_stoch_Fisher_KPP} in fact does \emph{not} correspond to the mesoscopic limit of a (spatially-extended/infinite dimensional) birth-death process~\citep{doering_interacting_2003} but is instead written down directly in analogy to the discrete trait Wright-Fisher diffusion 

\begin{linenomath*}
	\begin{equation}
		dp = p(1-p)dt + \sqrt{p(1-p)}dW_t
	\end{equation}
\end{linenomath*}

While Eq.~\ref{classical_stoch_Fisher_KPP} can be connected with birth-death processes via duality~\citep{doering_interacting_2003} and can be derived as a scaling limit of either a contact process~\citep{mueller_stochastic_1995} or a spatial $\Lambda$-Fleming-Viot process~\citep{barton_modelling_2013}, I show here that it does not arise as the expected mesoscopic limit of a measure-valued \emph{birth-death} process. Instead, the mesoscopic limit I obtain is Eq.\ref{example_stoch_Fisher_KPP}, which is also the equation expected from the more rigorous measure-theoretic approach to studying the processes I consider in this paper (see section 4.2 in~\cite{champagnat_unifying_2006} with $\eta = 1$).

\subsubsection*{The infinite population limit}
In the infinite population limit, we see from Eq. \ref{deterministic_traj} that the density field obeys the PDE
\begin{linenomath*}
	\begin{equation}
		\frac{\partial \phi}{\partial t} = \phi\left(1-\phi\right) + \frac{\sigma^2_m}{2}\nabla^2_x\phi
	\end{equation}
\end{linenomath*}
This is the famous Fisher-KPP equation and is widely used to model spatial evolution~\citep{perthame_concentration_2007,berestycki_non-local_2009,barton_modelling_2013}.

\phantomsection
\addcontentsline{toc}{section}{References cited in the Supplementary Information}
\printbibliography[title=References cited in the Supplementary Information]
	
\stopcontents[TOC]
\end{refsection}

\end{document}